\documentclass[usegraphicx,useAMS,usenatbib]{mn2e}
\pdfoutput=1
\usepackage[T1]{fontenc}
\usepackage{verbatim}
\usepackage{amsfonts,amsmath,amssymb} 
\usepackage{times}
\usepackage[italic]{mathastext} 
\usepackage{enumitem}
\usepackage[colorlinks,linkcolor=blue,citecolor=blue,urlcolor=blue]{hyperref}
\usepackage{color}
\usepackage{upgreek}
\usepackage{flushend}

\newcommand{\wtg}{WtG}
\newcommand{\clash}{CLASH}
\newcommand{\cccp}{CCCP}
\newcommand{\locuss}{LoCuSS}

\newcommand{\rockstar}{{\tt Rockstar}}

\newcommand{\gammat}{\mbox{$\gamma^{\intercal}$}}

\newcommand{\pzs}{\mbox{$p_\mathrm{phot}(z_{\mathrm{s},i})$ }}
\newcommand{\zsi}{\mbox{$z_{\mathrm{s},i}$}}
\newcommand{\zlj}{\mbox{$z_{\mathrm{l},j}$}}
\newcommand{\etan}{\mbox{$e^{\intercal}$}}
\newcommand{\etanij}{\mbox{$e^{\intercal}_{j,i}$}}

\newcommand{\deltasig}{$\Delta \Sigma$}
\newcommand{\dsig}{$\Delta\Sigma$}
\newcommand{\dsigr}{$\Delta\Sigma(R)$}

\newcommand{\scinv}{$\Sigma^{-1}_{\mathrm{crit}}$}
\newcommand{\hinv}{h^{-1}}

\newcommand{\photoz}{\mbox{photo-$z$}}
\newcommand{\photozs}{\mbox{photo-$z$s}}

\newcommand{\redmapper}{redMaPPer}

\newcommand{\ngmix}{\textsc{ngmix}}

\newcommand{\annz}{\textsc{annz2}}
\newcommand{\skynet}{\textsc{skynet}}
\newcommand{\tpz}{\textsc{tpz}}
\newcommand{\bpz}{\textsc{bpz}}
\newcommand{\balrog}{\textsc{Balrog}}

\newcommand{\neffngmix}{5.7}
\newcommand{\ngmixlowzbiasneg}{$-0.04$}

\newcommand{\njack}{40}
\newcommand{\shapenoise}{0.22}
\newcommand{\rmax}{30}
\newcommand{\imshape}{{\textsc{im3shape}}}
\newcommand{\snr}{S/N}

\newcommand{\msun}{\mathrm{M}_{\odot}}

\newcommand{\lfsens}{\mbox{\boldmath $s$}}

\newcommand\avg[1]{\langle #1 \rangle}
\newcommand{\Scrit}{\Sigma_{\rm crit}}
\newcommand{\Mpc}{\mathrm{Mpc}}
\newcommand{\hMpc}{h^{-1}\ \Mpc}

\newcommand{\fcl}{f_\mathrm{cl}}

\newcommand{\Mobs}{M_{\rm obs}}
\newcommand{\Mtrue}{M_{\rm true}}

\newcommand{\Rmis}{R_{\rm mis}}
\newcommand{\fmis}{f_{\rm mis}}
\newcommand{\cmis}{c_{\rm mis}}

\newcommand{\lkhd}{{\cal L}}
\newcommand{\calB}{{\cal B}}
\newcommand{\calM}{{\cal M}}
\newcommand{\calC}{{\cal C}}

\usepackage{eso-pic}

\AddToShipoutPictureBG*{%
  \AtPageUpperLeft{%
    \hspace{0.9\textwidth}%
    \raisebox{-1in}{%
      \makebox[0pt][l]{\textnormal{DES 2015-0148}}
}}}%

\AddToShipoutPictureBG*{%
  \AtPageUpperLeft{%
    \hspace{0.9\textwidth}%
    \raisebox{-1.2in}{%
      \makebox[0pt][l]{\textnormal{FERMILAB-PUB-16-468-AE}}
}}}%

\title[DES WL mass calibration of redMaPPer clusters]{Weak-lensing mass calibration of redMaPPer galaxy clusters\\ in Dark Energy Survey Science Verification data}

\author[Melchior, Gruen, McClintock, Varga, Sheldon, Rozo et al.]{%
\parbox{\textwidth}{\Large\raggedright%
P.~Melchior$^1$\thanks{corresponding author: \href{mailto:peter.melchior@princeton.edu}{peter.melchior@princeton.edu}},
D.~Gruen$^{2,3}$\thanks{corresponding author: \href{mailto:dgruen@stanford.edu}{dgruen@stanford.edu} (Einstein Fellow)},
T.~McClintock$^{4}$,
T. N. ~Varga$^{5,6}$,
E.~Sheldon$^{7}$,
E.~Rozo$^{4}$,
A.~Amara$^{8}$,
M.~R.~Becker$^{9,2}$,
B.~A.~Benson$^{10,11}$,
A.~Bermeo$^{12}$,
S.~L.~Bridle$^{13}$,
J.~Clampitt$^{14}$,
J.~P.~Dietrich$^{15,16}$,
W.~G.~Hartley$^{17,8}$,
D.~Hollowood$^{18}$,
B.~Jain$^{14}$,
M.~Jarvis$^{14}$,
T.~Jeltema$^{18}$,
T.~Kacprzak$^{8}$,
N.~MacCrann$^{13}$,
E.~S.~Rykoff$^{2,3}$,
A.~Saro$^{16}$,
E.~Suchyta$^{19}$,
M.~A.~Troxel$^{20,21}$,
J.~Zuntz$^{13}$,
C.~Bonnett$^{22}$,
A.~A.~Plazas$^{23}$,
T.~M.~C.~Abbott$^{24}$,
F.~B.~Abdalla$^{17,25}$,
J.~Annis$^{10}$,
A.~Benoit-L{\'e}vy$^{26,17,27}$,
G.~M.~Bernstein$^{14}$,
E.~Bertin$^{26,27}$,
D.~Brooks$^{17}$,
E.~Buckley-Geer$^{10}$,
A. Carnero Rosell$^{28,29}$,
M.~Carrasco~Kind$^{30,31}$,
J.~Carretero$^{32,22}$,
C.~E.~Cunha$^{2}$,
C.~B.~D'Andrea$^{33,34}$,
L.~N.~da Costa$^{28,29}$,
S.~Desai$^{35}$,
T.~F.~Eifler$^{23}$,
B.~Flaugher$^{10}$,
P.~Fosalba$^{32}$,
J.~Garc\'ia-Bellido$^{36}$,
E.~Gaztanaga$^{32}$,
D.~W.~Gerdes$^{37}$,
R.~A.~Gruendl$^{30,31}$,
J.~Gschwend$^{28,29}$,
G.~Gutierrez$^{10}$,
K.~Honscheid$^{20,21}$,
D.~J.~James$^{38,24}$,
D.~Kirk$^{17}$,
E.~Krause$^{2}$,
K.~Kuehn$^{39}$,
N.~Kuropatkin$^{10}$,
O.~Lahav$^{17}$,
M.~Lima$^{40,28}$,
M.~A.~G.~Maia$^{28,29}$,
M.~March$^{14}$,
P.~Martini$^{20,41}$,
F.~Menanteau$^{30,31}$,
C.~J.~Miller$^{42,37}$,
R.~Miquel$^{43,22}$,
J.~J.~Mohr$^{15,16,5}$,
R.~C.~Nichol$^{33}$,
R.~Ogando$^{28,29}$,
A.~K.~Romer$^{12}$,
E.~Sanchez$^{44}$,
V.~Scarpine$^{10}$,
I.~Sevilla-Noarbe$^{44}$,
R.~C.~Smith$^{24}$,
M.~Soares-Santos$^{10}$,
F.~Sobreira$^{28,45}$,
M.~E.~C.~Swanson$^{31}$,
G.~Tarle$^{37}$,
D.~Thomas$^{33}$,
A.~R.~Walker$^{24}$,
J.~Weller$^{15,5,6}$,
Y.~Zhang$^{10}$
\begin{center} (The DES Collaboration) \end{center}
\small{\emph{Author affiliations are listed at the end of this paper.}}
}
}

\begin{document}
\date{}
\pagerange{\pageref{firstpage}--\pageref{lastpage}} \pubyear{2016}
\maketitle
\label{firstpage}

\begin{abstract}
   We use weak-lensing shear measurements to determine the
    mean mass of optically selected galaxy clusters
    in Dark Energy Survey Science Verification data.
    In a blinded analysis, we split the sample of more than 8,000 \redmapper\
    clusters into 15 subsets, spanning ranges in the richness parameter $5 \leq \lambda \leq
    180$ and redshift $0.2 \leq z \leq 0.8$, and fit the averaged mass density
    contrast profiles with a model that accounts for seven distinct sources of
    systematic uncertainty: shear measurement and photometric redshift errors;
    cluster-member contamination; miscentering; deviations from the NFW halo
    profile; halo triaxiality; and line-of-sight projections.  We combine the
    inferred cluster masses to estimate the joint scaling relation between mass,
    richness and redshift, 
    $\calM(\lambda,z) \varpropto M_0 \lambda^{F} (1+z)^{G}$.
    We find $M_0 \equiv \avg{M_{200\mathrm{m}}\,|\,\lambda=30,z=0.5}=\left[ 2.35 \pm
    0.22\ \rm{(stat)} \pm 0.12\ \rm{(sys)} \right] \cdot 10^{14}\ \msun$,
    with $F = 1.12\,\pm\,0.20\ \rm{(stat)}\, \pm\, 0.06\
    \rm{(sys)}$ and $G = 0.18\,\pm\, 0.75\ \rm{(stat)}\, \pm\, 0.24\
    \rm{(sys)}$.  The amplitude of the mass--richness relation 
    is in excellent agreement with the weak-lensing calibration of \redmapper\ clusters in 
    SDSS by \citet{Simet2016} and with the \citet{Saro2015} calibration based
    on abundance matching of SPT-detected clusters.
    Our results extend the redshift range over which the
    mass--richness relation of \redmapper\ clusters 
    has been calibrated with weak lensing from $z\leq
    0.3$ to $z\leq0.8$.  Calibration uncertainties of shear measurements and
    photometric redshift estimates dominate our systematic error budget and
    require substantial improvements for forthcoming studies.
 \end{abstract}

\begin{keywords}
  cosmology: observations,
  gravitational lensing: weak,
  galaxies: clusters: general
\end{keywords}

\section{Introduction} \label{sec:intro}

The growth of massive structures and the cosmic expansion rate depend
directly on the energy constituents of the Universe and the behavior of gravity
at a range of scales.  In the currently favored cosmological model, the energy
density at the present epoch is dominated by dark matter and dark energy,
with spacetime evolving according to the standard theory of gravity, General Relativity.
Within this model, the number of halos of a given mass and the history of cosmic expansion
depend sensitively on the relative amount and detailed properties of both dark
matter and dark energy.
Alternative theories of gravity may make
different predictions for the number density of halos at a given cosmic
epoch for the same expansion history \citep[see e.g.][]{schmidtetal09}.
Thus, much can be learned about cosmology,
and potentially about gravity itself, by studying the abundance of massive
structures as a function of their mass and redshift \citep[][]{Allen2011,weinbergetal13}.

Clusters of galaxies are thought to directly correspond to the largest dark matter halos,
the number density of which is particularly sensitive to the dark energy.
However, clusters are typically identified not by the total mass of their halo,
but by a related observable.  Thus, cosmological inference from cluster abundance
requires a cluster catalog with measurements of some
observable with a well understood selection function, a theoretical prediction
for the abundance of halos as a function of mass and redshift for different
cosmologies, and, crucially, a mass-observable relation (MOR) that connects the
observable and true mass of a halo.

The abundance of dark matter halos is, in principle, predictable for a given
cosmological model purely from $N$-body simulations
\citep[e.g.][]{Tinker2008,Bocquet2016}.  However, most cluster observables that
are readily measured in large surveys are a manifestation of complex
astrophysical processes.  Some common observables are counts of galaxies above
a threshold luminosity \citep{KoesterCatalog07,Rykoff2014}, X--ray emission
from hot gas \citep{Piffaretti2010,Mehrtens2012}, and Compton scattering of
cosmic microwave background photons off electrons in that same gas
\citep{Hasselfield2013,PlanckXXVII2015,Bleem2015}.  Predicting the
corresponding MOR for these observables from first principles, or from
simulations, is not straightforward.  
At this time, it is more practical to determine the relationship empirically based on an observable proxy of the cluster mass.
These empirical measurements are similarly challenging; 
uncertainties in the MOR have in fact become the main impediment to
reliable inference of cosmological parameters from cluster abundances
\citep[e.g.][]{RozoCosmo09,PlanckXXIV2015,Bocquet2015,Mantz2015}.

Currently, the conceptually cleanest method for calibrating cluster masses is
gravitational lensing, the deflection of light from
background objects due to all matter contained in a foreground cluster
\citep[e.g.][]{Johnston07,Gruen2014,vonderLinden2014,Hoekstra2015,Simet2015,Okabe2015,vanUitert2016}.
In this work we will use lensing to measure the mass of clusters found
in the Dark Energy Survey \citep[DES,][]{DES05.1,DES15.1}.

The Dark Energy Survey   is a 5,000 square degree survey of the southern sky
using the 4-meter Blanco Telescope and the Dark Energy
Camera \citep{Flaugher2015} at the Cerro Tololo Inter-American Observatory.
The primary goal of the survey is to constrain
the distribution of dark matter in the Universe, and the amount and properties
of dark energy, including its equation of state.  Due to the large area,
depth, and image quality of DES,
the data will support optical identification of a large number ($\approx$~100,000) of
galaxy clusters and groups up to a redshift $z\approx 1$.
The potential for probing cosmology with these optically selected clusters can
be realized only if the MOR is well understood.

In this paper, we present the first ensemble, or ``stacked``, lensing measurements of optically
selected clusters from DES.  We use the redMaPPer cluster catalog
\citep{Rykoff2016}, generated from the Science Verification data taken before
the first official DES observing season.  For the lensing measurements, we use
catalogs of lensed galaxies constructed from the same data
\citep{Jarvis2016}.  

The goal of this work is twofold. First, we measure the
statistical relationship between the number of galaxies in a cluster and the underlying
halo mass -- i.e. the MOR -- and compare our findings to results in the literature. 
Second, we develop and test a new analysis pipeline, which we
will apply to the much larger DES data sets currently being acquired and
processed. We especially seek to fully account for systematic effects, considering biases in shear and photometric redshift measurements, cluster member contamination in the lensed background source sample, miscentering and triaxiality of clusters, projection of multiple clusters along the line of sight, and deviations of halo profiles from the analytical form.

The structure of this paper is as follows. In \autoref{sec:algo}, we introduce
the data used in this work. In \autoref{sec:stackedDS} we describe our
methodology for obtaining ensemble cluster density profiles from stacked 
weak-lensing shear measurements. A comprehensive set of tests and corrections 
for systematic effects is presented in \autoref{sec:systematics}.
The model of the lensing data and the inferred stacked cluster masses are given
in \autoref{sec:modeling}.
The main result, the mass--richness--redshift relation of \redmapper\ clusters in DES,
is presented in \autoref{sec:mass_richness_relation}.
We compare our results to other published works
in the literature in \autoref{sec:comparisons} and conclude in \autoref{sec:summary}.

For the purpose of this analysis, we assume a flat $\Lambda$CDM cosmology with
$\Omega_{\rm m}=0.3$ and $H_0=70$ km s$^{-1}$ Mpc$^{-1}$. Distances and masses,
unless otherwise noted, are defined as physical quantities with this choice of
cosmology, rather than in comoving coordinates. We denote the mass inside
spheres around the cluster center as $M_{200\rm{m}}$, corresponding to an
overdensity factor of 200 with respect to the mean matter density at the
cluster redshift.

\section{The DES Science Verification data}
\label{sec:algo}

\begin{figure}
  \includegraphics[width=\linewidth]{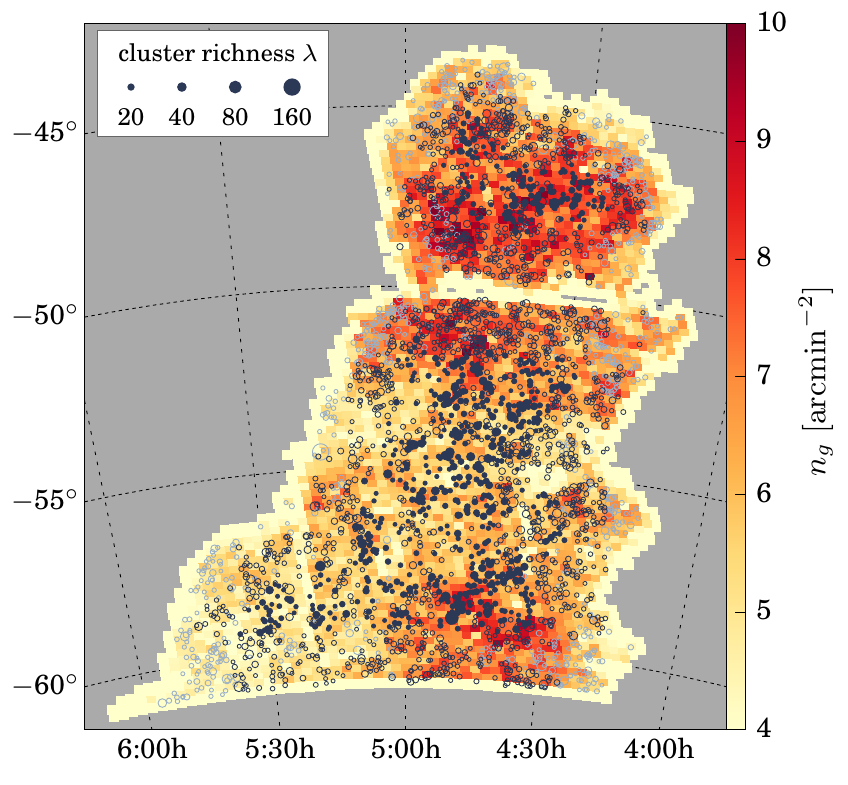}
  \caption{The distribution of \redmapper{} clusters (\autoref{sec:redmapper})
  in the DES SV footprint,  overlaid on the density of galaxies used for the weak-lensing analysis
    (\autoref{sec:shearcat}). The size of the cluster markers is scaled by the richness $\lambda$, i.e. the number of member galaxies.
    Clusters that are surrounded by galaxies over the entire radial range
    of 30 Mpc are shown as filled circles (other clusters are denoted by open circles).
    For the purpose of clarity, clusters with $\lambda < 10$ are omitted.}
\label{fig:footprint}
\end{figure}

DES began taking official survey data in August 2013 \citep{Diehl14.1}.
Before this, a small Science Verification (SV) survey was conducted from
November 2012 to February 2013. For this work, we restricted our measurements to the largest
contiguous portion of the SV area, dubbed ``SPT-East" and shown in \autoref{fig:footprint},
an area of approximately 139 square degrees in the eastern part of the region
observed by the 2500-square-degree South Pole Telescope Sunyaev-Zel'dovich Survey \citep{Carlstrom11.1}.
We briefly introduce the three main data products utilized in this
work and refer to the respective publications for details.
We want to emphasize that all catalogs used
in this work have already been made public%
\footnote{\url{http://des.ncsa.illinois.edu/releases/sva1}}
to facilitate external review and further scientific exploration.

\subsection{Cluster catalog}
\label{sec:redmapper}

\begin{figure}
\includegraphics[width=\linewidth]{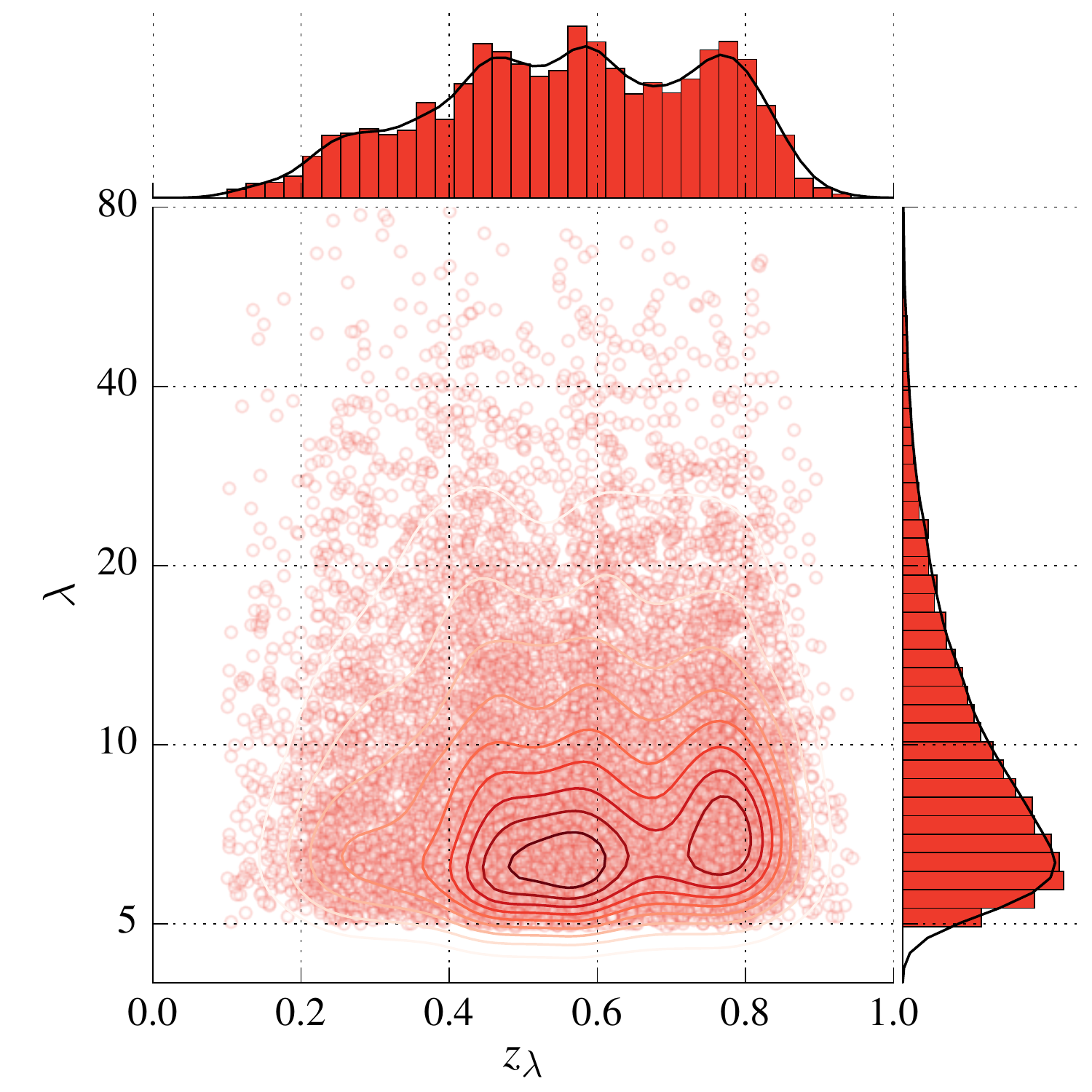}
\caption{Redshift--richness distribution of \redmapper\ clusters in the DES SV catalog,
  overlaid with density contours to highlight the densest regions. At the top and on the right
  are histograms of the projected quantities, $z_\lambda$ and $\lambda$, respectively, with
  smooth kernel density estimates (\emph{black lines}).}
\label{fig:z_lambda}
\end{figure}

We rely on photometrically selected clusters identified in the DES SV data set
using the \redmapper\ cluster finding algorithm \citep{Rykoff2014}.  The resulting
cluster catalog was presented in \citet{Rykoff2016}, and the distribution in the footprint is
shown in \autoref{fig:footprint}.  We note that \redmapper\ has
had multiple public releases corresponding to different versions of the algorithm.
The catalog we employ corresponds to version 6.3.3 of the algorithm.

Briefly, \redmapper\ identifies galaxy clusters as overdensities of red-sequence galaxies.
It iteratively trains a model of the red-sequence as a function of redshift, and utilizes
this model to assign a cluster membership probability to every galaxy in the vicinity of a
cluster.  The cluster richness $\lambda$ is the sum of the membership probabilities of the
galaxies within a cluster radius $R_\lambda$. The radius scales with richness as
$R_\lambda = 1.0(\lambda/100)^{0.2}h^{-1}$ Mpc, a choice that was found to minimize 
the scatter in the MOR \citep{Rykoff12}. 
Because of the interdependence of $\lambda$ and $R_\lambda$, richness estimation involves finding the self-consistent pair of these parameters (cf. \citealt{RozoEstimator08}, their section 3.2).

Cluster central galaxies are chosen using a probabilistic approach that weights not just galaxy luminosity,
but also local galaxy density, and demands consistency between the photometric redshift of the
central galaxy and the cluster redshift. The mean of the \redmapper\ centering probabilities is 0.81, i.e.
we expect 81\% of the clusters to be properly centered, which in good agreement with estimates
from XMM-Newton, Chandra, and SPT cluster detections \citep[][their section 6.2.4]{Rykoff2016}.

High-redshift clusters can be identified only in the deepest survey regions, so the
redshift range probed by the catalog varies from location to location.
Specifically, at any given location, the maximum redshift $z_{\rm max}$
is set by requiring that the survey depth at that location be sufficient to ensure
a 10-$\sigma$ detection of \redmapper\ member galaxies brighter than $0.2\, L_*$ 
for a cluster at redshift $z_{\rm max}$.

The distribution of richness and redshift for the cluster sample is shown in \autoref{fig:z_lambda}. 
\citet{Rykoff2016} show that the photometric redshift performance of \redmapper\ on the DES SV region 
is $\sigma_{z_\lambda}\big/(1+z)\lesssim 0.01$, while the abundance and redshift evolution of the sample suggests 
that a richness $\lambda=20$ corresponds to a halo mass
$M_{500\mathrm{c}} \approx 10^{14}\ \msun$ or 
$M_{200\mathrm{m}} \approx 1.8\times 10^{14}\ \msun$,
with a variation of mass with richness expected to be roughly linear.

\subsection{Shear catalog}
\label{sec:shearcat}

We use the official DES shear catalogs presented in \citep{Jarvis2016}.  Two
separate catalogs were created, \ngmix\ \citep{Sheldon2014} and \imshape\
\citep{Zuntz13}.  These catalogs were found to be consistent when selection
effects were taken into account \citep{Jarvis2016}.  Both catalogs are adequate
for the purposes of estimating shear correlation functions, including
tangential shear analyses such as the one pursued in this work.  Here, we adopt
the \ngmix\ catalog, which has a higher effective number density (\neffngmix\
per square arcmin.) of sources because it combines the image data from three
bands ($r,i,z$) instead of relying on any single band.

The \ngmix\ catalog was generated using an implementation of the ``lensfit''
algorithm \citep{Miller07}, a galaxy model fitting technique.  An exponential
model was fit to each galaxy image, and the full likelihood surface of
the model parameter space was explored.  The ellipticity statistic $\boldsymbol{e}$ was taken to
be the mean $\langle \boldsymbol{e}_\mu \rangle$ across this surface, where
$\mu$ denotes the two ellipticity components.

During exploration of the likelihood surface, a centrally concentrated prior
was applied to the ellipticity.  This prior stabilizes and limits the
exploration of the likelihood.  However, application of an ellipticity prior
results in a shear calibration error: a naive sum over the ellipticities does
not recover the true applied shear for low \snr\ measurements.  Essentially,
the sensitivity of the ellipticity as a shear estimator is reduced due to
suppressing high ellipticity regions of the parameter space.  A correction
for this calibration bias was calculated, called the ``sensitivity'' \lfsens\
\citep[their eq. 7.10]{Jarvis2016}, which yields a shear estimator of the following form
\begin{equation}
  \label{eq:sensitvity}
  \tilde{\boldsymbol{g}}_\mu = \frac{ \sum_i \boldsymbol{e}_{i,\mu} }{\sum_i \lfsens_{i,\mu}},
\end{equation}
for all sources $i$ that experienced the same constant shear. It is important
to note that for any single galaxy \lfsens\ is a very noisy quantity. The sensitivity correction is
therefore applied only for an entire source ensemble.
Also, while \lfsens\ has two components, it does not
transform as a polarization.  For tangential shear measurements, we rotate the
galaxy shapes into the tangential frame around each cluster, but this cannot be
done for the sensitivities.  Instead, we simply average the two sensitivity
components into one,
\begin{equation}
\label{eq:s_avg}
  s_i = \frac{1}{2}(\lfsens_{i,1} + \lfsens_{i,2}).
\end{equation}
We have verified in simulations that using the mean sensitivity gives
equivalent results to using the individual components when recovering
a constant shear.  This reflects the fact that the two sensitivity
components are equal in the mean, within our uncertainties.

\subsubsection*{Blinding}

During the development of the analysis presented here, the \ngmix\ shear catalog
was blinded by an unknown factor between 0.9 and 1.0 \citep[their section 7.5]{Jarvis2016}.
Only after the data and modeling pipeline presented here had successfully passed all internal tests 
were the unblinded shears processed.

However, after unblinding we uncovered an inconsistency in the interpretation of radial bin limits between data and model.
We note that this discovery was made by directly assessing whether the model properly approximates the data,
and not by comparing our mass richness relation to literature results or expectations. 
We repeated the blinded analysis with the corrected pipeline, checked the results for consistency again, 
and only then continued with the unblinded analysis.

\subsection{Photometric redshift catalog}
\label{sec:photo-z}

Inference of physical quantities such as mass from a lensing signal requires
knowledge of the redshift distribution of the source galaxies being lensed.
DES has explored a broad range of photometric redshift estimators \citep{Sanchez2014}. Four of the best
performing codes were selected for a detailed characterization of the impact of photometric redshift uncertainties
on weak lensing studies such as this one \citep{BonnettPhotoz2015}.
These were \annz\ \citep{Sadeh2015,Collister2004}, \skynet\ \citep{Graff2013}, \tpz\
\citep{Carrasco2013}, and \bpz\ \citep{Benitez2000, Coe2006}.
The first three of these are machine learning codes, while \bpz\ is a template-based method.
All four methods noted above were used to produce a full probability
distribution $p(z)$ for every source galaxy in the DES shear catalogs, finding  comparable performances.

We note that \citet{BonnettPhotoz2015} found it necessary to shift the probability distribution
recovered from \bpz\ upwards by 0.05 in order to counteract intrinsic biases
in \bpz\ that arise due to limitations in the template set employed.
Since our analysis relies on the same photometric data and photometric redshift outputs
as \citet{BonnettPhotoz2015},
we follow these authors in applying a systematics shift of 0.05 to the photometric
redshift distributions from \bpz\ for all following analyses, and in 
using the \skynet\ \photozs\ for the fiducial calculations in this work.
We present a detailed characterization
of photometric redshift uncertainties in \autoref{sec:sysphotoz}.

\section{Stacked lensing measurements}
\label{sec:stackedDS}

\begin{figure*}
  \includegraphics[width=\textwidth]{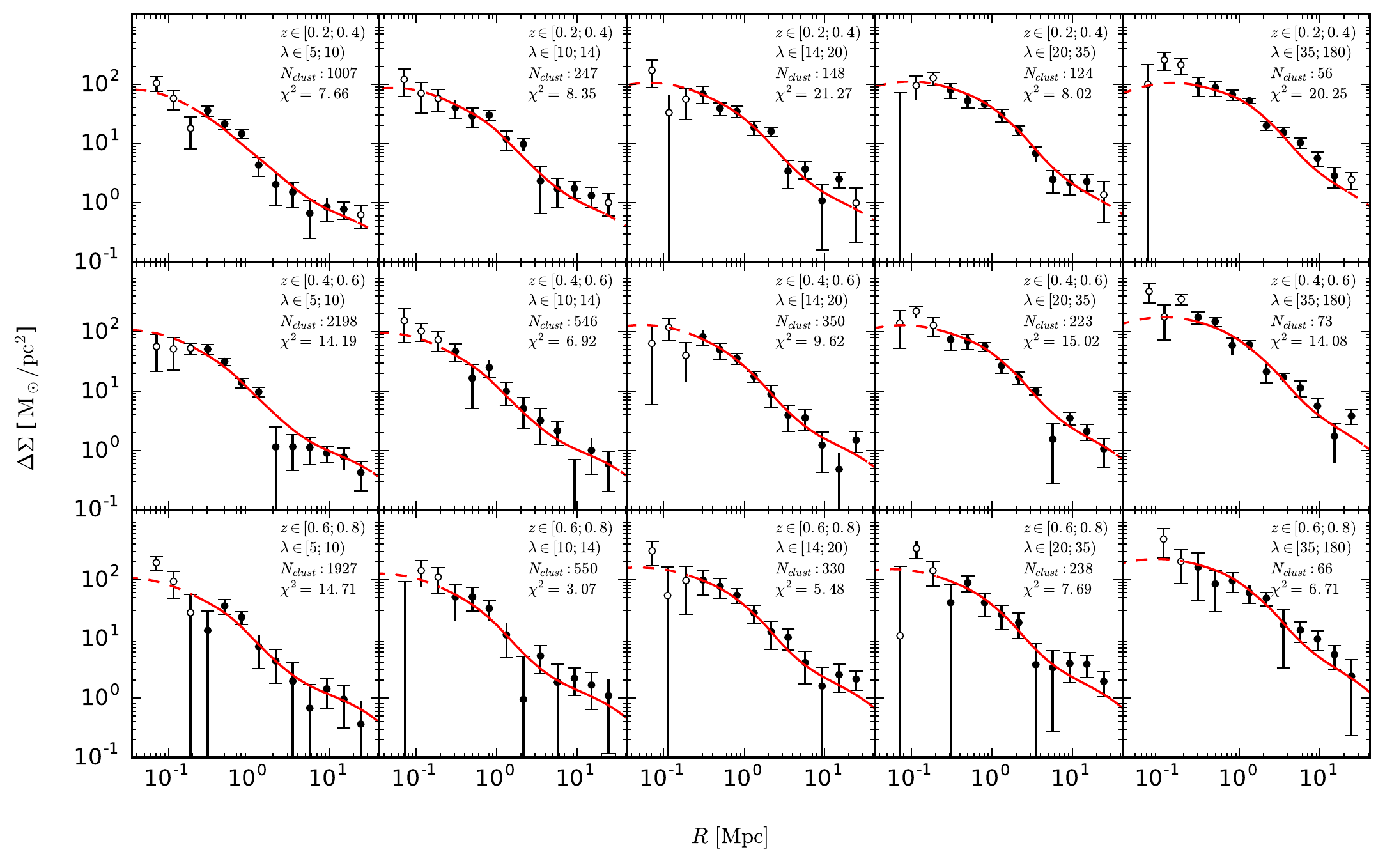}
  \caption{Mean $\widetilde{\Delta\Sigma}$ for cluster subsets split in redshift
    $z_\mathrm{l}$ (\emph{increasing from top to bottom}) and $\lambda$
      (\emph{increasing from left to right}) with errors from jackknife resampling
      (see \autoref{sec:covariance}). Our best-fit model (\emph{red curve}) includes
      dilution from cluster member galaxies (\autoref{sec:boost_factors}) and
      miscentering (\autoref{sec:centering}); see \autoref{fig:best_fit_example} for details.
      Data points considered unreliable and 
      therefore excluded from further analysis (below 200 kpc or above 1 deg) are indicated
      by open symbols and dashed lines. The profiles and jackknife errors are calculated
      after the subtraction of the random-point shear signal (see \autoref{sec:randoms}).}
    \label{fig:DeltaSigma}
\end{figure*}

We briefly summarize the methodology for stacked cluster lensing measurements,  and
refer the interested reader to \citet{Bartelmann01.1}  and \citet{Sheldon04.1} for details.

\autoref{eq:sensitvity} defines an estimator for
the ``reduced shear''
\begin{equation}
\label{eq:reduced_shear}
  \boldsymbol{g} \equiv \frac{\bgamma}{1-\kappa},
\end{equation}
a non-linear combination of the gravitational shear $\bgamma$ and convergence $\kappa$.
\emph{Weak} lensing is characterized by $\boldsymbol{g} \approx \bgamma$, which
is an accurate approximation for this study: we excise areas close to
cluster centers to avoid high shear and difficulties of photometry and shape measurements in crowded
regions (cf. \autoref{sec:boost_factors} and \autoref{sec:results}).

The weak shear by a foreground mass concentration induces correlations in the
shapes of background galaxies, such that, on average, galaxies images are stretched
tangentially with respect to the center of mass.
The mean tangential shear \gammat\ at a
distance $R$ from a cluster is related to the surface mass density $\Sigma$ of the cluster via
\begin{equation}
\label{eq:DeltaSigma}
  \Delta\Sigma \equiv \overline{\Sigma}(<R) - \overline{\Sigma}(R) = \Sigma_{\rm crit}\ \gammat (R)\mathrm{,}
\end{equation}
where  $\Sigma(x,y) = \int_z \rho(x,y,z)$ is the line-of-sight projection of
the physical mass density $\rho$, $\overline{\Sigma}(<R)$  is the mean surface
density within $R$, and $\overline{\Sigma}(R)$ is the azimuthally averaged surface
density at radius $R$.  We will take the source
ellipticity, rotated to the tangential frame, as a noisy estimator
\begin{equation}
    \gammat \approx \etan + \mathrm{noise},
\end{equation}
where the noise is due to both the intrinsic ellipticities of the
source galaxies and measurement noise.

The strength of the lensing signal imprinted on background galaxies is
modulated by the critical surface density
\begin{equation}
  \label{eq:sigmacrit}
  \Sigma_{\rm crit}(z_\mathrm{s}, z_\mathrm{l}) = \frac{c^2}{4\uppi G}\frac{D_\mathrm{s}}{D_\mathrm{l} D_\mathrm{ls}}\mathrm{,}
\end{equation}
where $D_\mathrm{s}$, $D_\mathrm{l}$, $D_\mathrm{ls}$ refer to the angular diameter distances to the
source, to the lens, and between lens and source, respectively.
\autoref{eq:sigmacrit} provides a mechanism to weight each lens-source pair to maximize the
signal-to-noise ratio of \dsig. For most lensing analyses (including this one), the
distances of the sources must be estimated from photometric rather
than spectroscopic redshifts, which demands the
introduction of an effective critical density for each lens-source pair
\begin{equation}
  \langle\Sigma_{\rm crit}^{-1}\rangle_{j,i} = \int\ \mathrm{d}\zsi\ \pzs \Sigma_{\rm crit}^{-1}(\zsi, \zlj)
  \label{eq:sigmacritinv}
\end{equation}
that averages over the redshift probability distribution \pzs\ of
source $i$, evaluated for lens $j$.%
\footnote{In this work we also estimate the
  cluster redshifts photometrically. However, the error on those estimates is of
  order $\Delta z\approx0.01$ \citep{Rykoff2016}, which is negligible compared to
  both the width of the lens redshift bins we adopt and of the source redshift
  distributions.  In what follows, we will therefore treat the cluster redshifts
  as exact.}
We can combine the previous equations in this section to arrive 
at an estimator $\widetilde{\Delta\Sigma}_{j,i}$ for
each lens-source pair $(j,i)$.  The optimal estimator for the 
stacked density profile $\Delta\Sigma$ is then
a weighted sum over these individual estimators, so that
\begin{equation}
  \widetilde{\Delta\Sigma} = \frac{\sum\limits_j^\mathrm{lens}\ \sum\limits_i^\mathrm{src} w_{j,i}\ \etanij\
  \big/ \langle\Sigma_{\rm crit}^{-1}\rangle_{j,i}}{\sum_{j,i} w_{j,i}}.
  \label{eq:delsig}
\end{equation}
where the weights
\begin{equation}
  w_{j,i} = \langle\Sigma_{\rm crit}^{-1}\rangle_{j,i}^2\ \big/\ \sigma^2_{\gamma,i}
  \label{eq:wdeltasigma}
\end{equation}
are chosen to minimize the variance of the resulting estimator
\citep{Sheldon04.1}.\footnote{Our weight depends on the measured ellipticity noise, which
correlates with the galaxy ellipticity, and thus can induce
a bias in the recovered shear.  However, in \citet{Jarvis2016} we
found that weighting effects are sub-dominant to other sources of bias for the
\ngmix\ catalog.}
The quantity $\sigma^2_\gamma$ is the uncertainty on the shear measurement,
and combines the uncertainty in the shear due to intrinsic galaxy
shapes $\sigma_{\mathrm{SN}}$ with the uncertainty in the
ellipticity measurement,
\begin{align}
    \sigma^2_{\gamma,i} = \sigma^2_{\mathrm{SN}} + \frac{1}{2} \left(\mathsf{C}_{11,i} + \mathsf{C}_{22,i}\right),
\end{align}
where $\mathsf{C}_{11} + \mathsf{C}_{22}$ is the trace of the ellipticity subset of the
covariance matrix produced by the \ngmix\ code.  We take $\sigma_{\mathrm{SN}}$ =
\shapenoise.
Although the use of this simple $\sigma^2_\gamma$ is
not optimal \citep{Bernstein02.1}, it only slightly increases the variance of \dsig.

Our particular choice of ellipticity measurement requires division by a mean
sensitivity to produce unbiased results (see \autoref{sec:shearcat}).  We therefore modify
our \deltasig\ estimator to use the mean weighted sensitivity,
\begin{equation}
  \label{eq:delta_sigma_est}
  \widetilde{\Delta\Sigma} = \frac{\sum\limits_j^\mathrm{lens}\ \sum\limits_i^\mathrm{src} w_{j,i}\ \etanij\ \big/ \langle\Sigma_{\rm crit}^{-1}\rangle_{j,i}}{\sum_{j,i} w_{j,i}\, s_i}.
\end{equation}
Efficient codes to compute the estimator in \autoref{eq:delta_sigma_est}
are publicly available. For this work, we used {\sc xshear}%
\footnote{\url{https://github.com/esheldon/xshear}} and cross-checked
its results with an independent implementation.%
\footnote{\url{https://github.com/pmelchior/shear-stacking}}
The resulting shear profiles are shown in \autoref{fig:DeltaSigma}.
Clusters were split into three subsets $z_\mathrm{l} = [0.2, 0.4)$, $[0.4, 0.6)$, $[0.6, 0.8)$
 and five subsets $\lambda = [5, 10)$, $[10, 14)$, $[14, 20)$, $[20, 35)$, $[35, 180)$
with 13 logarithmically spaced radial bins between 0.05 and 30 Mpc.
The redshift splitting allows for constraints on a possible redshift evolution of the MOR,
while removing only
a small number of clusters at $z_\mathrm{l} > 0.8$ whose lensing weights are low
because their redshift is larger than that of most galaxies in the shear catalog.
The richness splitting was adjusted to provide roughly equal signal-to-noise ratio for
each cluster subset. The number of radial bins was set so that all of them are
populated, and the outer cutoff corresponds to the spatial size of the jackknife
regions in our covariance estimation scheme.

\subsection{Data covariance matrices}
\label{sec:covariance}

The \dsigr\ profiles we measure from our data deviate from the true
mean \dsigr\ profiles of clusters in a given richness-redshift subset by a
statistical uncertainty that we characterize by a covariance matrix $\mathsf{C}_{\widetilde{\Delta\Sigma}}$.
Contributions to $\mathsf{C}_{\widetilde{\Delta\Sigma}}$ include shape
noise, uncorrelated large-scale structure along the line of sight
\citep[e.g.][]{Hoekstra01a,Hoekstra03,Hoekstra2011,Umetsu2011}, and intrinsic
variations of cluster profiles at fixed mass
\citep[e.g.][]{Metzler01,Gruen2011,Becker2011,Gruen2015}.

For a stacked cluster lensing analysis in a common footprint, a single source
galaxy may be within the maximum search radius of multiple galaxy clusters.
Thus the cluster \dsig\ measurements are not fully independent, and the
covariance matrix will have significant off-diagonal terms, particularly
on large scales.
This is
exacerbated by the spatial co-location of clusters and background galaxies
(cf. \autoref{fig:footprint}) due to a combination of variations in
survey depth, which affect the detectability of galaxies and \redmapper\ clusters similarly,
and cluster and galaxy clustering.

For estimating $\mathsf{C}_{\widetilde{\Delta\Sigma}}$, we therefore use a spatial
jackknife scheme designed to account for the covariance of the
measurements, which we expect to depend on scale. To this end, we split the source sample into $K=\njack$ simply-connected regions $\mathcal{R}_k$ by
running a $k$-means algorithm on the sphere.%
\footnote{\url{https://github.com/esheldon/kmeans_radec/}}
For each such $\mathcal{R}_k$ and each richness-redshift subset, we calculate \autoref{eq:delta_sigma_est} for all
lenses $j\notin \mathcal{R}_k$ and denote it $\widetilde{\Delta\Sigma}_{(k)}$.
We then calculate the covariance matrix according to \citet{Efron82.1},
\begin{equation}
\label{eq:jackknife}
  \mathsf{C}_{\widetilde{\Delta\Sigma}} = \frac{K-1}{K} \sum_k^K (\widetilde{\Delta\Sigma}_{(k)} - \widetilde{\Delta\Sigma}_{(\cdot)})^T \cdot (\widetilde{\Delta\Sigma}_{(k)} - \widetilde{\Delta\Sigma}_{(\cdot)})\mathrm{,}
\end{equation}
where $\widetilde{\Delta\Sigma}_{(\cdot)} = \tfrac{1}{K}\sum_k \widetilde{\Delta\Sigma}_{(k)}$.
We note that in the above expression, $\widetilde{\Delta\Sigma}$ concatenates all redshift
and richness subsets into a single data vector, enabling us to detect covariance across all redshift, richness, and radial bins.

\begin{figure}
 \includegraphics[width=\linewidth]{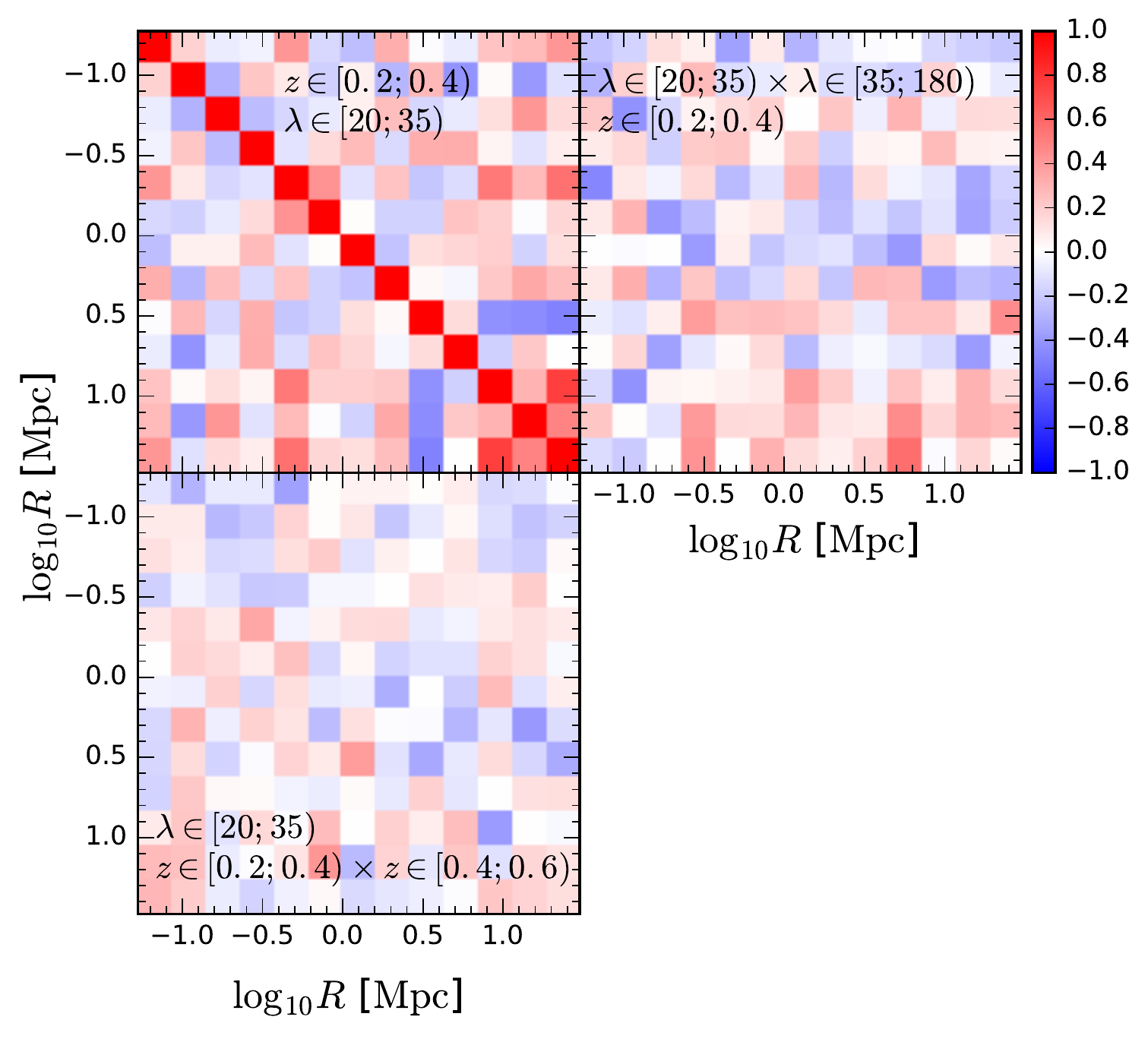}
 \caption{Correlation matrix of $\widetilde{\Delta\Sigma}$ of a single
   richness-redshift subset with $z\in[0.2,0.4)$ and $\lambda \in [20,35)$
   for \njack\ jackknife regions \emph{(upper left panel)}.
   Off-diagonal blocks show the correlation matrix between the lensing profile of the reference
   subset and the lensing profiles of the neighboring redshift subset
   ($z\in[0.4,0.6)$, \emph{lower left}),
   and the neighboring richness subset ($\lambda\in[35,180)$, \emph{upper right}).}
 \label{fig:cmatrix}
\end{figure}

The upper left panel of \autoref{fig:cmatrix} shows the resulting correlation
matrix for one reference cluster subset. On smaller scales the diagonal is
dominant, but---as expected---off-diagonal terms are present for the largest scales.
We also test for cross-correlations between the profiles measured for clusters from
different richness and redshift subsets,
and find them to be small (cf. upper right and lower left panel of
\autoref{fig:cmatrix}).
We therefore will make the assumption of no cross-correlation between different cluster
subsets in our likelihood analysis presented in \autoref{sec:complete_likelihood}.
Future analyses with larger DES data sets will result in significantly reduced uncertainties of the covariance matrices, allowing us to properly include any correlation that may have remained undetected here because of the modest size of the SV data set.

In addition, we also performed covariance estimation with smaller jackknife
regions ($K=100$) and cluster-by-cluster jackknifing, yielding similar results on
small scales. We note that jackknife schemes are prone to underestimation of
the covariance on scales that exceed the size of the jackknife patches
\citep[e.g.][for the case of shear auto-correlations]{Friedrich2016}. In the case
of $K=40$ and DES SV, this corresponds to angular scales of approximately one degree.
For this reason, we exclude \deltasig\ measurements at $R >1$ deg from further
analysis (cf. \autoref{fig:DeltaSigma}).

The covariance matrix estimated from the data is noisy, and its inverse is therefore a biased estimate of the true inverse covariance \citep[e.g.][]{Kaufman67,Hartlap07}. In addition, the use of noisy covariances leads to additional uncertainty in estimated parameters \citep{Dodelson2013}. The size of these effect depends on the details of the jackknife scheme, data vectors, and structure of the true covariance.

To calibrate the effect, we generate random realizations of 10 uncorrelated zero-mean Gaussian random variables in 40 jackknife patches. This is close to the jackknife scheme and true covariance of the data vectors used in \autoref{sec:results} for the estimation of masses. For each realization, we estimate the covariance matrix from \autoref{eq:jackknife}, and invert it to find the best fit mean and its uncertainty. We compare the estimated uncertainty to the actual scatter of the best fit over a large number of realizations.

We find that the actual uncertainty of the best fit is $\approx30\%$ larger than the one estimated with the inverse jackknife covariance. Approximately half of this excess uncertainty is corrected when applying the de-biasing factor of eqn. 17 in \citet{Hartlap07} to the inverse covariance, the remaining half is consistent with the expectations from \citet{Dodelson2013}.
We correct both effects by rescaling the statistical uncertainty of mass estimates in \autoref{sec:results} by a factor of 1.3. Future work would benefit from the use of less noisy (e.g. simulation-based) or analytical \citep{Gruen2015} covariance matrices.

\section{Systematics} \label{sec:systematics}

\subsection{Shear systematics}
\label{sec:sysshear}

The \ngmix\ catalog passed an extensive set of null tests on real data and simulations
\citep{Jarvis2016}, which we will briefly summarize here. We adopt the bias
parameterization
\begin{equation}
\label{eq:shear_bias}
  \tilde{\boldsymbol{g}} = (1 + m)\boldsymbol{g} + \alpha\boldsymbol{e}_\mathrm{PSF} + c.
\end{equation}
Note that $m$, $\alpha$, and $c$ are in principle two-component variables, but in
our experience the values in both components are identical within
errors.  We will therefore
treat them as scalars.  PSF size modeling errors resulted in a multiplicative
bias $|m_{\rm PSF}|<0.01$,
while inaccurate deconvolution led to PSF leakage $\alpha < 0.01$.
Both of these effects are negligible compared to the remaining systematics
described in this section, and are therefore ignored in our systematic error budget.

Lacking an absolute shear calibration source, we cannot test the overall shear
calibration using real data.  We instead adopted a set of simulations
\citep{Jarvis2016} based on the real galaxy images from the COSMOS imaging data
\citep{lilly07}.  The selection of these COSMOS galaxies does not perfectly
match our selection in DES data; the COSMOS field is quite small and thus
subject to cosmic variance.  Thus we cannot infer any detailed information
about the shear bias in DES data; the bias in real data may be more or less
than the bias we see in the simulations.  Without additional information, however, we
choose to model our systematics based on what we found in these simulations.

It is important to note that the \imshape\ catalog was re-calibrated directly from
these simulations to minimize biases caused by pixel noise, whereas \ngmix\ did not require that step.
For \ngmix, the multiplicative shear calibration error $m$ was seen to be
consistent with zero for galaxies at redshift $z\approx1$, but as large as
\ngmixlowzbiasneg\ for sources at $z\approx0.3$.  
We believe that shear inference for the lower redshift sources in the COSMOS galaxy sample
is primarily affected by ``model bias'', introduced by fitting an exponential
disk model to a galaxy population that has a large number of bulge-like
galaxies. 
We note that the bulk of the cluster sample is at redshift $z_\lambda > 0.5$ (cf.
\autoref{fig:z_lambda}), so that these sources, already a minority in the
shape catalogs, receive significant weight only for the lowest redshift clusters.
The simulation results therefore suggests that the multiplicative shear bias is
controlled to $|m|\leq0.03$.

However, a detailed comparison of the \ngmix\ and \imshape\ shape catalogs
performed by \citet{Jarvis2016} found the residual systematic uncertainties to be
larger: $|m|\leq 0.05$. This finding may reflect the differences between the
simulations from which \imshape\ was calibrated, and the DES SV data, for instance 
in the redshift-dependent bulge fraction \citep{Jarvis2016}.
Rather than adopting a top-hat prior with $|m| \leq 0.05$, we use a Gaussian
prior with the same variance.  Doing so preserves the total error budget, while
avoiding an inappropriate sharp cut on $|m|$.  The corresponding Gaussian prior
(to be used in the likelihood analysis of \autoref{sec:complete_likelihood}) is
$m=0.00 \pm 0.03$, where the error is to be interpreted as standard deviation.

Additive errors were found to be below the cosmic-shear requirement of $|c| <
0.002$ for the SV survey area \citep{Jarvis2016}.  Below we perform additional null
tests, particularly useful for stacked cluster lensing analyses.

\subsubsection{B-modes and quadrant checks} 
\label{sec:bmodes}

We also perform an additional test for residuals by projecting the galaxy
ellipticities onto the direction $45^\circ$ off the tangent. This so-called ``B-mode''
should be zero in the mean if the signal we measure is solely from
gravitational lensing.
Using the jackknife covariance, we test for significant deviation of the $\chi^2$
of B-mode shears from zero. The stacked B-mode signal of clusters with $\lambda>20$
in all redshift bins combined yields a reduced $\chi^2/N_{\rm dof}\approx16/10$ for
10 radial bins, consistent with no systematic at the $p>0.1$ level.

Additive errors of the $c$- or $\alpha$-type are cancelled in \gammat\ when sources separated by $90^\circ$ are averaged,
which we implicitly exploit whenever sources are evenly distributed around each lens.
But survey boundaries, as well as holes due to masking of bright stars and other
features, result in a non-uniform distribution of source galaxy positions. The
DES SV footprint is rather small, so our large maximum radius $R_\mathrm{max} = \rmax\ $Mpc
around each cluster lens often intersects at least one such boundary.
We identify clusters for which that is the case with the so-called ``quadrant check'' \citep[][their section 3.2.4]{Sheldon04.1} and show them with open circles in \autoref{fig:footprint}.

\begin{figure}
\includegraphics[width=\linewidth]{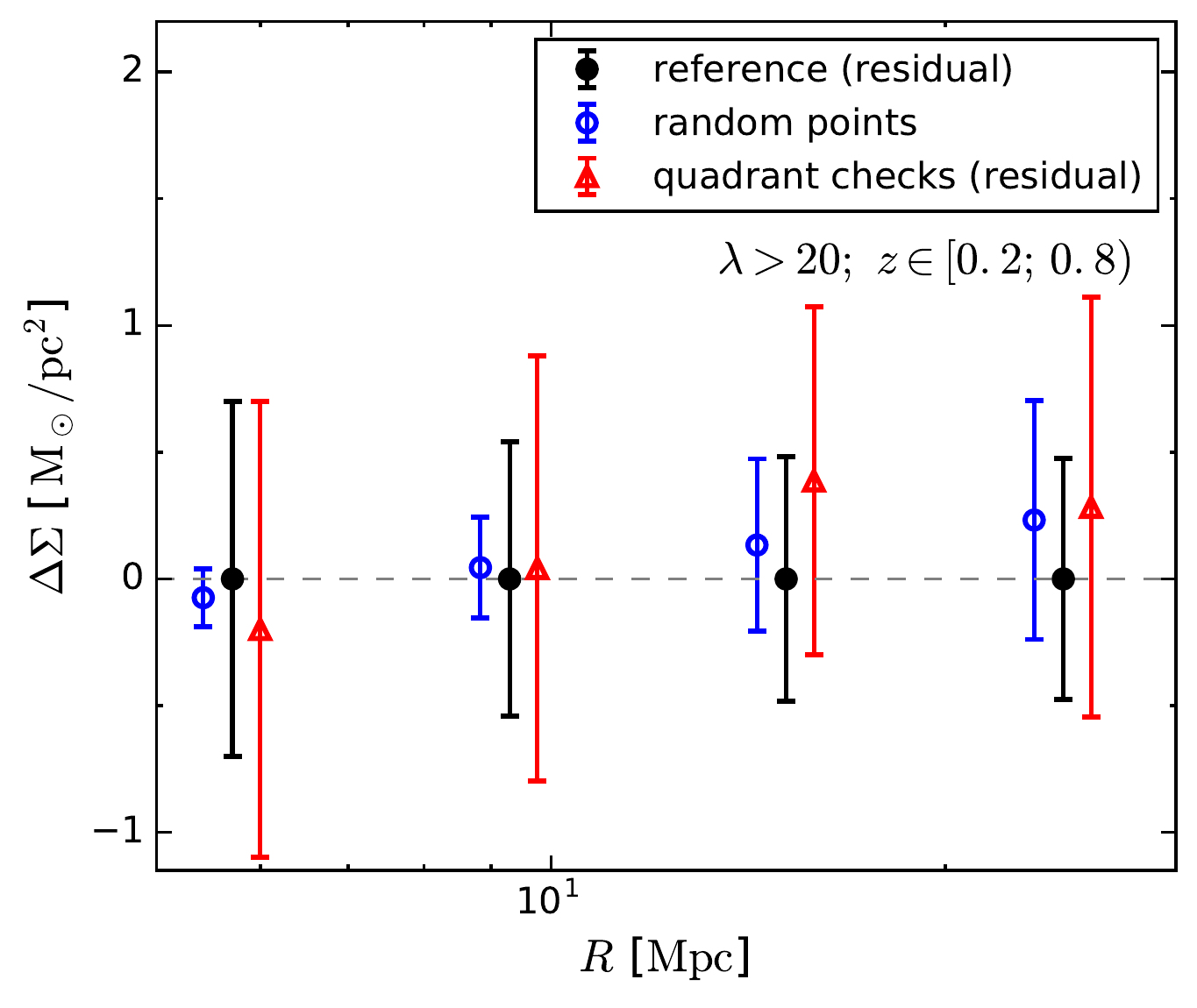}
 \caption{Corrections for additive shear systematics: the shear signal around random
   points, drawn such as to mimic the distribution of \redmapper\ clusters in the DES SV
   footprint (\emph{blue open circles}, cf. \autoref{sec:randoms}); the difference
   in the shear signal caused by rejecting the clusters that are excluded after the quadrant
   check (\emph{red triangles}, cf. \autoref{sec:bmodes}); for comparison, the
   errors of the \deltasig\  profile after random-point subtraction (\emph{black circles})}
 \label{fig:rand_quad}
\end{figure}

We show the difference between the $\widetilde{\Delta\Sigma}$ profiles before and after
the quadrant-based rejection in \autoref{fig:rand_quad} (red markers). We can
see that the difference is largest at scales outside of 10 Mpc, but
not statistically significant. Requiring that clusters pass
the quadrant check substantially increases statistical uncertainties on
large scales, where many source-lens pairs get rejected. As it
does not appear to be beneficial in this work, we will therefore not demand that
clusters need to pass the check and will utilize the entire cluster sample for further analysis.

\subsubsection{Random point test} \label{sec:randoms}

Despite being not significant in the test of \autoref{sec:bmodes},
additive shear systematics may still be present on all scales, which can
lead to small spurious shear signals in our analysis. A simple correction
can be made by measuring the tangential shear around a set of random points,
which reproduce the redshift and richness distribution of clusters in the DES
SV footprint, and subtracting it from the actual cluster signal.

To this end, we use the weighted random point sample as described in
\citet[][their section 3.6]{Rykoff2016}, generated from the survey mask and
redMaPPer maximum redshift maps. Since the effective survey geometry varies
with cluster redshift and richness, we split these random points in the same
way as our cluster sample and measure $\Delta\Sigma$ profiles around each subset.

\autoref{fig:rand_quad} shows results from the random point shear measurement
as blue markers. We find no significant random shears even on the largest scales.
To benefit from a correction of potential shear systematics below the detection
limit, we will use random-point subtracted shear profiles for the rest of our analysis.

\subsection{Correction for cluster members in the shear catalog} \label{sec:boost_factors}

\begin{figure*}
\includegraphics[width=0.48\textwidth]{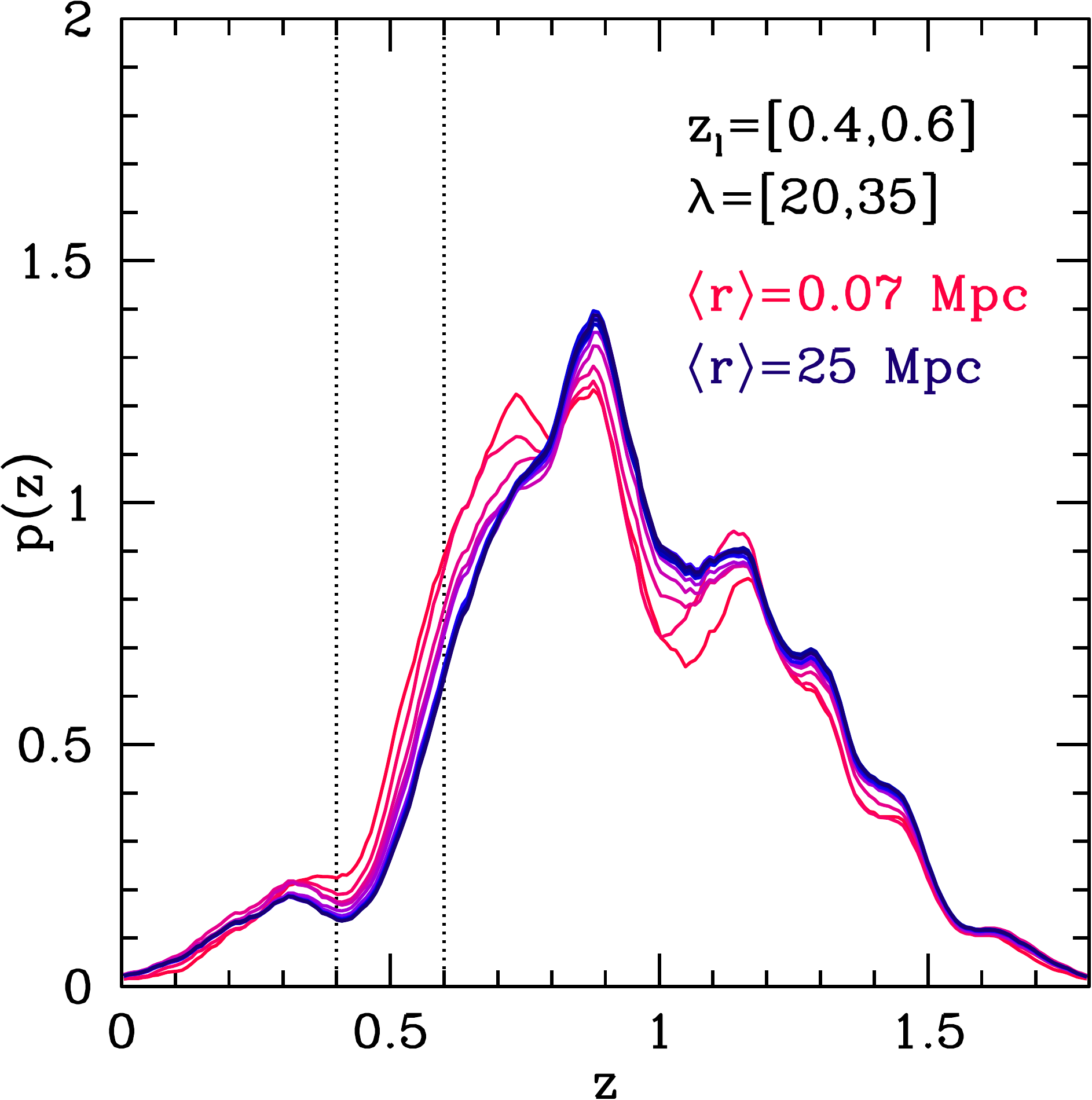}
\hspace{10pt}
\includegraphics[width=0.48\textwidth]{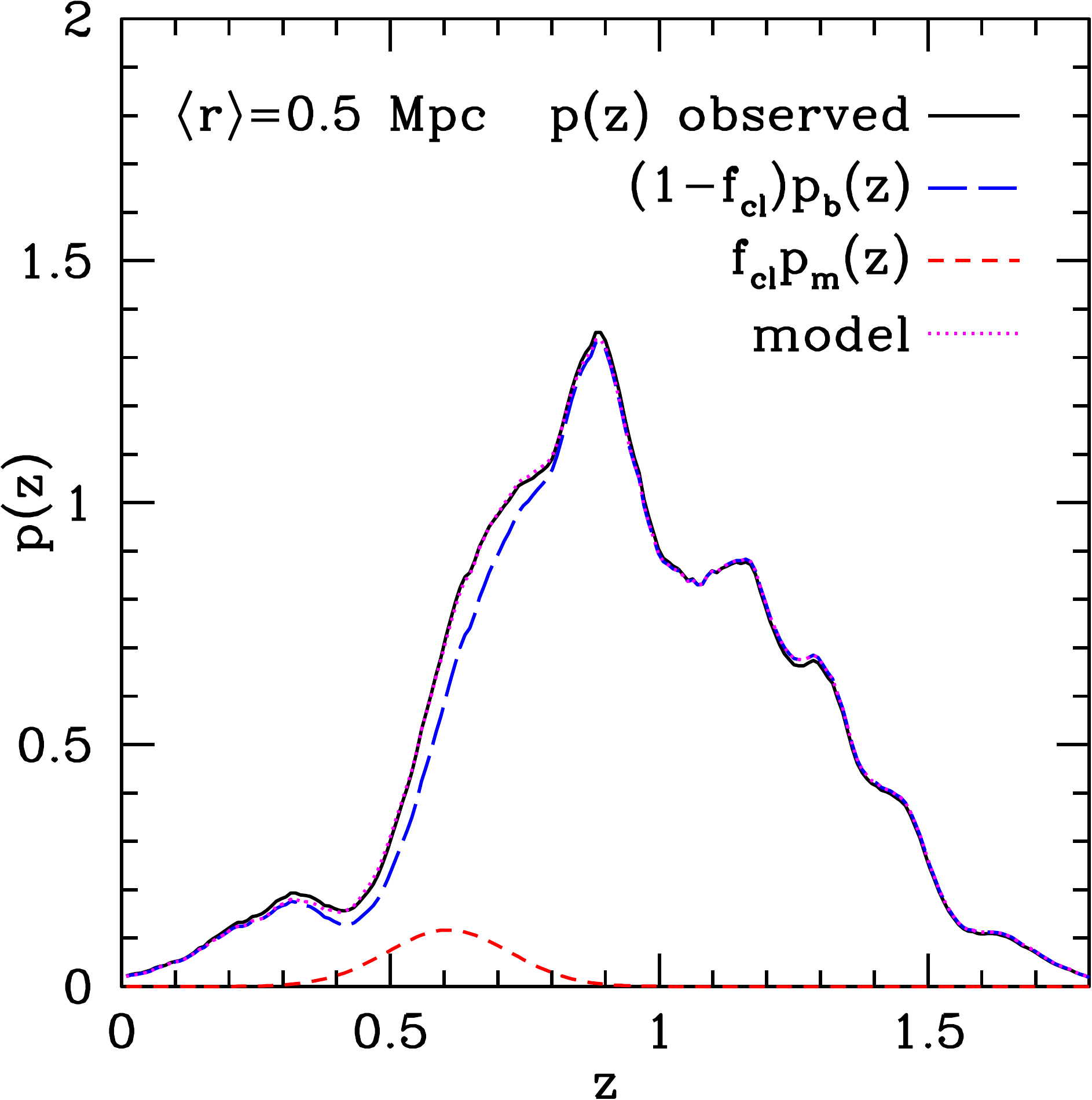}
\caption{Estimation of cluster member contamination of the source sample by $p(z)$
  decomposition. Left-hand panel: stacked, lensing-weighted $p(z)$ of sources
  around clusters in one redshift-richness bin from the innermost (red) to outermost
  (blue) distance bin. Right-hand panel: decomposition of observed $p(z)$ in one radial bin
  (black, solid line) into the sum (magenta, dotted line) of field
  distribution $p_b(z)$ as measured around random points (blue, long-dashed line)
  and Gaussian $p_m(z)$ of cluster members (red, short-dashed line).}
\label{fig:fcl}
\end{figure*}

Due to photometric uncertainties and ambiguity of the available color
information, the estimated $p(z)$ of the source galaxies is quite broad, and a
source at lower redshift than the cluster will get some non-zero weight in the
analysis.  This is properly accounted for when calculating $\langle\Sigma_{\rm
crit}^{-1}\rangle_{j,i}$ of a source-lens pair in \autoref{eq:sigmacritinv}.
However, cluster members must be treated specially, as their redshift
distribution is essentially a delta function centered on the cluster redshift.
This delta function is not properly accounted for in the $p(z)$, and
thus not included in the integral to calculate
\scinv. Each member gets a non-zero weight, but adds zero to the mean shear.%
\footnote{\label{fn:ia}This is only true in absence of intrinsic alignments.
Observational results on mild radial alignment of cluster members towards
the halo center vary \citep{Hao2011,Sifon2015}. If such alignment is present,
it would lead to $\langle\Delta\Sigma_{j,i}\rangle_m < 0$. We consider it a
higher-order term and will thus neglect it in this work.}
Furthermore, cluster members are highly concentrated near the cluster center,
so there is a strong radially dependent bias in the inferred \dsig.

As a first step for alleviating this effect, we exclude all galaxies that are
likely cluster members from the shape catalog. We run the redMaPPer
algorithm for member identification \citep{Rozo2015} and reject all galaxies consistent with
being a cluster member down to the magnitude limit of the survey and out to $1.5 R_{\lambda}$.

While this significantly reduces the contamination from cluster members, a fraction
of them remains in the source catalog. We estimate their contribution to the signal,
the so-called \emph{boost factors} \citep[e.g.][their section 4.1]{Sheldon04.1},
for each cluster subset as a function of $R$, and correct the negative impact in \autoref{sec:boost_factor_model}.
We make use of a method similar to the one described in
\citet[][their section 3.1.3]{Gruen2014},%
\footnote{We found the standard method based on correlation functions
\citep{Sheldon04.1} unreliable due to the small footprint of the DES SV data set.}
which method is based on the decomposition of the redshift distribution of
source galaxies into a field galaxy component and a cluster member component.

Consider all lens-source pairs $(j,i)$ in a cluster richness-redshift subset and
at some projected separation $R$. Each source-lens pair yields an estimate of
$\Delta\Sigma_{j,i}=\etanij\big/\langle\Sigma_{\rm crit}^{-1}\rangle_{j,i}$ that enters
the mean measured $\widetilde{\Delta\Sigma}$ with relative weight $w_{j,i}$
(cf. \autoref{eq:wdeltasigma} and \autoref{eq:delta_sigma_est}).
Assuming that we can split the source-lens pairs into pairs with field galaxies $b$ and pairs
with cluster member galaxies $m$, we expand
\begin{equation}
\widetilde{\Delta\Sigma}=\frac{\sum_{j,i\in b}w_{j,i}\Delta\Sigma_{j,i}+\sum_{j,i\in m}w_{j,i}\Delta\Sigma_{j,i}}{\sum_{j,i\in b}w_{j,i}+\sum_{j,i\in m}w_{j,i}} \; .
\end{equation}
Taking the expectation value and using that $\langle\Delta\Sigma_{j,i}\rangle=\Delta\Sigma$
for field galaxies and $\langle\Delta\Sigma_{j,i}\rangle = 0$ for cluster
members,\textsuperscript{\ref{fn:ia}} we find
\begin{equation}
\langle\widetilde{\Delta\Sigma}\rangle=\frac{\sum_{j,i\in b}w_{j,i}}{\sum_{j,i\in b}w_{j,i}+\sum_{j,i\in m}w_{j,i}}\Delta\Sigma = (1-f_{\rm cl})\Delta\Sigma\;\mathrm{,}
\end{equation}
where we defined the fractional weight of cluster member galaxies
\begin{equation}
f_{\rm cl}(R)=\frac{\sum_{j,i\in m}w_{j,i}}{\sum_{j,i}w_{j,i}}=1-\frac{\sum_{j,i\in b}w_{j,i}}{\sum_{j,i}w_{j,i}}
\end{equation}
and the radial dependence stems from the selection of pairs $(j,i)$ for both $m$ and $b$.
In principle, this allows us to correct the data for the effect of member dilution via
\begin{equation}
\label{eq:DeltaSigma_boosted}
\widetilde{\Delta\Sigma}_{\rm corr}(R)=\frac{\widetilde{\Delta\Sigma}(R)}{1-f_{\rm cl}(R)} \; .
\end{equation}
In practice, rather than correct the data for this effect,
we choose to dilute the predicted lensing signal to match the observational data when
modeling the recovered $\Delta\Sigma$ profiles.

The remaining task is to measure $f_{\rm cl}$ from the weighted, estimated redshift distribution of sources,
\begin{equation}
p(z)=\frac{\sum_{j,i}w_{j,i}p_i(z)}{\sum_{j,i}w_{j,i}} \; .
\end{equation}
The left-hand panel of \autoref{fig:fcl} shows the $p(z)$ of sources in a set
of annuli around the cluster centers.
We observe a systematic increase of low redshift sources as one moves from large cluster-centric
distances to small radii, clearly illustrating the effects of cluster membership contamination of
the source galaxy catalog.

We decompose the observed photometric redshift distribution $p(z)$ as a weighted sum
of a field galaxy and a member galaxy component,
\begin{align}
p(z)&=\frac{\sum_{j,i\in b}w_{j,i}p_i(z)+\sum_{j,i\in m}w_{j,i}p_i(z)}{\sum_{j,i}w_{j,i}} \nonumber \\
&=(1-f_{\rm cl})\ p_b(z)+f_{\rm cl}\ p_m(z) \; ,
\end{align}
using the stacked, weighted redshift distributions $p_b(z)$
and $p_m(z)$ of the respective samples.

We measure $p_b(z)$ from the redshift distribution of source galaxies
around random points. To ensure sampling from the same distribution of survey
depth as around the actual lenses, we bin and weight
the random-source pairs in the the same way as the lens-source pairs.
The weighted cluster member redshift distribution is not well constrained by the data in each individual bin. 
We can, however, find robust constraints on its mean and variance with a joint fit of a
Gaussian with a common mean and width to all radial
bins of a cluster (sub)sample, and a free amplitude in each radial
bin that corresponds to $f_{\rm cl}$. 

The right panel of \autoref{fig:fcl} shows the model
for one example radial bin at $\langle R \rangle = 0.5$ Mpc.
We can see that the model recovers an excess contribution (red dashed curve)
from objects that are not present in the field galaxy population. We note
that it is necessary to leave the mean value
of the Gaussian as a free parameter: due to the skewness of the lensing weight applied to the
stacked $p(z)$, which are zero at $z_\mathrm{s} \leq z_\mathrm{l}$ and positive at
higher source redshift, and the redshift prior for galaxies found in our survey,
the mean of the recovered Gaussian is not expected to coincide with the cluster redshift.
We have ensured that this method of decomposition yields consistent results
with (a) a decomposition with free mean and width of the Gaussian in each radial
bin (instead of fixed over all radial bins) and (b) a non-parametric
measurement of $(1-f_{\rm cl})$ by the ratio of integrals over $p(z)$ and $p_b(z)$
over $|z-z_{\rm cl}|>0.3$.

There is an implicit assumption inherent to this method, namely that the observed $p(z)$ deviate from
$p_b(z)$ only because of cluster-member contamination. This assumption may be violated for at least two reasons.
First, lensing magnification changes the redshift 
distribution of field galaxies in a complex way that depends on the details of galaxy types, 
redshifts and luminosity functions characterizing the lensing source sample, and
a part of this could in principle mimic cluster-member contamination. 
\citet{Gruen16} investigated this effect and find the bias on $\Sigma_{\rm crit}^{-1}$ for a DES-like source population to be below the percent level.

Second, \photoz\ estimates may be affected by measurement biases caused by the presence of clusters,
in particular an increased probability of blended sources in the cluster core.
To investigate this possibility, we compared the results of the $p(z)$-decomposition 
with an extension of the method employed by \citet[their section 3.4.1]{Melchior15.1}:
we make use of the \balrog\ catalog of fake objects in the DES SV footprint
\citep{Suchyta16.1} to compare the lensing weights of actual DES galaxies around
clusters with those of fake galaxies. When \balrog\ galaxies are matched to have
the same properties as the galaxies in our lensed source sample, we can infer the
effects of increased blended or light contamination in the dense cluster environments.
Due to the lack of a \photoz\ catalog for \balrog\ objects, we could only perform
the matching to DES galaxies in the shear catalog via proxies, for which we used
the SExtractor parameters \verb|FLUX_RADIUS| and \verb|MAG_AUTO| in the $i$ band.
Despite this limitation, we found good agreement of the boost factors obtained from
this method with the $p(z)$-decomposition presented above for the two high-redshift
subsamples and a 1-$\sigma$ discrepancy for the lowest cluster redshift subsample,
when restricted to $R \geq 200$ kpc.
We will revisit this issue in forthcoming analyses with larger cluster samples, and for this work
adopt the boost factors from the $p(z)$ decomposition but limit the acceptable
range of the shear profiles to $R\geq 200$ kpc.

\subsection{Photometric redshift systematics}
\label{sec:sysphotoz}

\citet{BonnettPhotoz2015} assessed the impact of photometric redshift systematics 
on weak lensing analyses of DES SV data, e.g. by characterizing how
$\avg{\Sigma_{\rm crit}^{-1}}$ varies as a function of lens redshift between the
various photometric redshift algorithms.

Here, we update the results of \citet{BonnettPhotoz2015} to account for the additional source weights that enter
into our estimator.  Specifically, consider the estimator for \autoref{eq:delsig},
\begin{equation}
\widetilde{\Delta\Sigma} = \frac{\sum_{j,i} w_{j,i}\ \etanij\ S_{j,i}^{-1}}{\sum_{j,i} w_{j,i}}\mathrm{,}
\end{equation}
where we have defined $S_{j,i} \equiv \Sigma_{{\rm crit},ji}^{-1}$ as the true expectation value of
$\Scrit^{-1}$ for the lens-source pair $(j,i)$.
The weights $w_{j,i}$ take the form $w_{j,i}=S_{j,i}^2\sigma_{j,i}^{-2}$.  The above
estimator is appropriate when one has spectroscopic redshifts for all lens-source pairs.

\begin{figure}
\includegraphics[width=\linewidth]{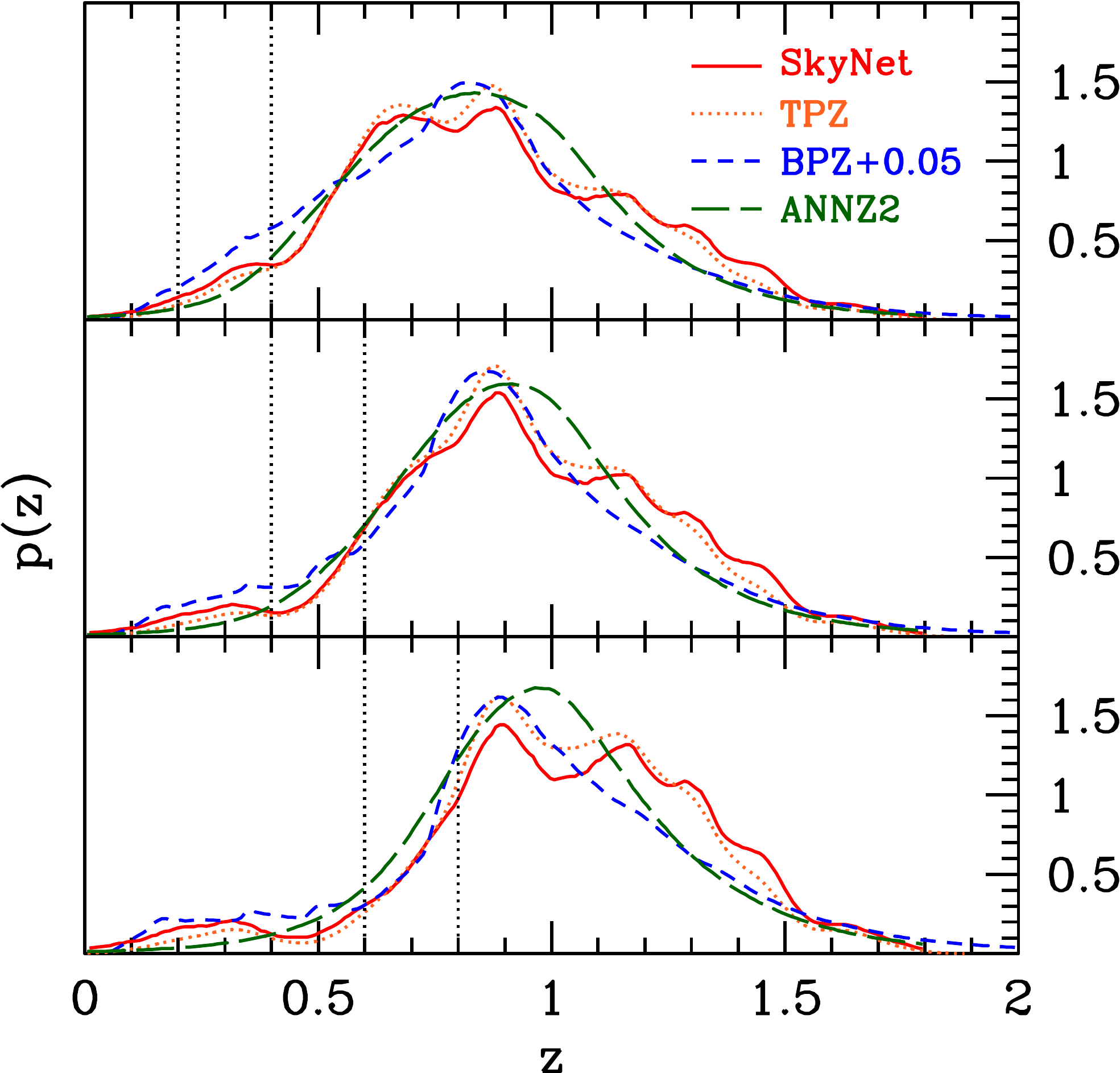}
\caption{Lensing-weighted, stacked $p(z)$ estimates for sources around
  clusters at $z\in[0.2,0.4]$ (\emph{top panel}, lens redshift range indicated by
  vertical dotted lines), $z\in[0.4,0.6]$ (\emph{central panel}) and $z\in[0.6,0.8]$
  (\emph{bottom panel}) from four different photo-z codes. The $p(z)$ were estimated
  from sampling sources around redMaPPer random points in the $\lambda>35$ and respective
  redshift subsets at a distance of $\approx1$~Mpc from the cluster, assigning the
  weight of \autoref{eq:wdeltasigma}. The distribution for \bpz\
  was shifted towards larger $z$ by 0.05.}
\label{fig:pzcomp}
\end{figure}

We wish to determine
how the estimator changes when we use a biased photometric redshift code for which
$S_{j,i}' = \avg{\Scrit^{-1}}_{j,i} \neq S_{j,i}$.
Defining $\epsilon_{j,i}$ via $S_{j,i}' \equiv S_{j,i}(1+\epsilon_{j,i})$, we have
\begin{equation}
\widetilde{\Delta\Sigma'} = \frac{\sum_{j,i} \sigma_{j,i}^{-2}\ S_{j,i}\ (1+\epsilon_{j,i})\ \etanij }{\sum_{j,i} \sigma_{j,i}^{-2}\ S_{j,i}^2\ (1+\epsilon_{j,i})^2}.
\end{equation}
We take the expectation value of the above equation, using $\avg{\etanij}= S_{j,i} \Delta\Sigma$.  Expanding
to first order in $\epsilon$, we arrive at
\begin{equation}
\label{eq:delta_DeltaSigma}
\avg{\widetilde{\Delta\Sigma'}} = \Delta\Sigma \left[ 1 - \delta \right]
\end{equation}
where we have defined
\begin{equation}
\label{eq:delta}
\delta \equiv \frac{ \sum_{j,i} w_{j,i}\epsilon_{j,i} }{\sum_{j,i} w_{j,i}}.
\end{equation}
The quantity $\delta$ has an alternative interpretation, as the difference of the true mean
inverse critical surface density $\avg{\Scrit^{-1}}$ from its estimate $\avg{\Scrit^{-1}}'$
based on a photometric redshift code,
\begin{equation}
\label{eq:alt_delta_DeltaSigma}
\avg{\Scrit^{-1}}' = \frac{ \sum_{j,i} w_{j,i}\ S_{j,i}' }{\sum_{j,i} w_{j,i} } = \avg{\Scrit^{-1}} \left[ 1 + \delta \right].
\end{equation}
In the absence of spectroscopic information, we adopt one of our photometric
redshift codes, \skynet, as the fiducial algorithm, and then estimate the relative
offset $\delta$ at fixed lensing weights $w_{j,i}$ from \autoref{eq:alt_delta_DeltaSigma}.
For precisely known lens redshifts one can write
\begin{equation}
\label{eq:thin-lens_sigma_crit1}
\begin{split}
\avg{\Scrit^{-1}}' &= \frac{1}{\sum_{j,i} w_{j,i}} \sum_{j,i} w_{j,i} \int \mathrm{d}z\ P(z)\ \Scrit^{-1}(z,z_{\rm lens}) \\
	&= \int \mathrm{d}z\ P_{\rm eff} (z)\ \Scrit^{-1}(z,z_{\rm lens})
\end{split}
\end{equation}
where we have defined the effective source redshift distribution
\begin{equation}
P_{\rm eff}(z) = \frac{\sum_{j,i} w_{j,i}\ P_i(z)}{\sum_{j,i} w_{j,i}}.
\end{equation}
The previous two equations provide a numerically convenient way to evaluate $\delta$.
Indeed, this is what was done in \citet{BonnettPhotoz2015}, albeit with unit weights
for all sources and a spectroscopic reference sample.

\begin{table}
\caption{Systematic difference $\delta$ of lensing weighted $\avg{\Scrit^{-1}}$ between different \photoz\ codes according to \autoref{eq:alt_delta_DeltaSigma} and \autoref{eq:thin-lens_sigma_crit1}. In the absence of spectroscopic redshifts for the source galaxies, we quote $\delta$ as the difference with respect to our reference pipeline \skynet. \bpz+0.05 refers to results obtained after shifting all reported redshifts upwards by 0.05 (cf. \autoref{sec:photo-z}). The statistical errors on $\avg{\Scrit^{-1}}$ are of order 0.01 and thus negligible compared to the systematic spread between methods.
}
\label{tab:photoz_table}
\centering
\begin{tabular}{llcc}
Lens sample & Code & $\avg{\Scrit^{-1}}^\prime$\ $[10^{-4} \msun^{-1} \mathrm{pc}^{2}]$ & $\delta$\ [\%]\\
\hline
$z\in[0.2,0.4]$ & \skynet & 2.94 & --- \\
& \annz & 2.96 & 0.7 \\
& \tpz  & 2.97 & 1.0 \\
& \bpz+0.05 & 2.80 & $-$4.8 \\
\hline
$z\in[0.4,0.6]$ & \skynet & 2.74 & --- \\
& \annz & 2.68 & 2.2 \\
& \tpz & 2.75 & 0.4\\
& \bpz+0.05 & 2.57 & $-$6.2 \\
\hline
$z\in[0.6,0.8]$ & \skynet & 2.16 & --- \\
& \annz & 1.98 & $-$8.3 \\
& \tpz & 2.16 & 0.0 \\
& \bpz+0.05 & 1.97 & $-$8.8 \\
\hline
\end{tabular}
\end{table}

We compute the effective source redshift distributions, averaging over the
lens redshifts of each of our three lens subsets, using the \redmapper\ random
points from the highest richness subset, in the radial bin around $R=1\, \hMpc$.
The results are shown in \autoref{fig:pzcomp}. Selecting a different
richness subset or radial aperture has no significant impact on these results.
\autoref{tab:photoz_table} lists the mean values of $\avg{\Scrit^{-1}}^\prime$ for
each of our \photoz\ codes and lens redshift bins, and the corresponding
$\delta$ values. As already observed by \citet{BonnettPhotoz2015}, the three
machine learning codes are in much better agreement with
each other than they are with \bpz, the only template-based code we consider,
even after applying the global upwards shift of 0.05 to account for limitations
in the template set (cf. \autoref{sec:photo-z}).

For the likelihood analysis in \autoref{sec:complete_likelihood}, we seek
to capture the spread and biases between the \photoz\ codes in the form of
a prior. One way to set such a prior would be to consider the range of $\delta$ values obtained from the different
algorithms, and then setting a flat top-hat prior over the range $\delta \in [\delta_{\rm min},\delta_{\rm max}]$.
However, we wish to allow for the possibility that $\delta$ is slightly beyond these limits by using a prior that
drops off smoothly.  To do so, we instead use a Gaussian prior that has the same mean and variance as the putative top-hat prior.
Our final Gaussian prior is therefore
\begin{equation}
\label{eq:delta_prior}
\begin{split}
\delta &= \frac{1}{2}\left(\delta_{\rm max}+\delta_{\rm min}\right) \pm \frac{0.577}{2}\left(\delta_{\rm max}-\delta_{\rm min}\right)\\
&= \begin{cases}
-0.019 \pm 0.017 & {\rm for}\  z\in[0.2,0.4] \\
-0.020 \pm 0.024 & {\rm for}\  z\in[0.4,0.6] \\
-0.044 \pm 0.025 & {\rm for}\  z\in[0.6,0.8].
\end{cases}
\end{split}
\end{equation}

Finally, we acknowledge that lensing magnification induces changes of the background source population that can cause additional biases specific to cluster lensing in the \photoz\ estimation. For a DES-like sample, resulting biases of $\Sigma_\mathrm{crit}^{-1}$ are of the order of 0.5\% and of opposite direction to the magnification-induced biases for the boost factors \citep{Gruen16}, so we ignore both for the remainder of this work.

\section{The stacked lensing signal}
\label{sec:modeling}

We seek to calibrate the mass--richness relation of \redmapper\ clusters.  We will therefore estimate
the mean cluster mass in each of our richness and redshift subsets by modeling the stacked
weak lensing signal of each cluster subset as if it originated from a hypothetical halo of mass $M$.
Below, we detail our model for the lensing signal as a function of the halo mass, and use numerical simulations
to calibrate how the recovered mass is related to the mean cluster mass.  We then consider how our model
needs to be extended to account for the systematics discussed in the previous sections,
as well as the additional complication caused by our limited knowledge of the true halo center.

\subsection{Surface density model}
\label{sec:surface_density_model}

\autoref{fig:DeltaSigma} shows the resulting estimates of the excess projected surface mass density
$\Delta\Sigma$ defined in \autoref{eq:DeltaSigma}. The quantities $\Sigma(R)$ and $\overline{\Sigma}(<R)$ are given by
\begin{equation}
  \label{eq:sigma_less_R}
  \overline{\Sigma}(<R) = \frac{2}{R^2}\int_0^R \mathrm{d}R^\prime \ R^\prime \ \overline{\Sigma}(R^\prime)
\end{equation}
and
\begin{equation}
  \label{eq:sigma_r}
  \overline{\Sigma}(R) = \int_{-\infty}^{+\infty} \mathrm{d}\chi\ \rho\left(\sqrt{R^2+\chi^2}\right)\mathrm{.}
\end{equation}
If the shear signal is caused by halos of mass $M$, the excess three-dimensional matter density is given by
\begin{equation}
\rho(r) = \rho_\mathrm{m}\ \xi_{\rm hm}(r\,|\,M)
\end{equation}
where $\rho_{\rm m} = \Omega_\mathrm{m}\,\rho_\mathrm{c}(1+z)^3$ is the
mean matter density in physical units at the redshift of the sample, $\rho_\mathrm{c}$ is the critical density, and
$\xi_\mathrm{hm}(r\,|\,M)$ is the halo--matter correlation function
evaluated for a halo of mass $M$.

We use the \citet{zuetal14} update to the \citet{Hayashi08} model of the halo--matter
correlation function.  Specifically, we set
\begin{equation}
  \label{eq:xi_hm}
  \xi_\mathrm{hm}(r\,|\,M) = \mathrm{max}\left\lbrace\xi_{\rm 1h}(r\,|\,M),\ \xi_{\rm 2h}(r\,|\,M)\right\rbrace\mathrm{,}
\end{equation}
where we have constructed the total $\xi_\mathrm{hm}$
from the so-called ``1-halo'' and ``2-halo'' terms.
We model the 1-halo term
as a \citet*[][hereafter NFW]{Navarro96.1} density profile $\rho_\mathrm{NFW}(r\,|\,M)$,
\begin{equation}
  \label{eq:1halo}
  \xi_{\rm 1h}(r\,|\,M) = \frac{\rho_\mathrm{NFW}(r\,|\,M)}{\rho_{\rm m}} - 1,
\end{equation}
with a statistical concentration--mass relation, for which we employ the \citet{DiemerKravtsov15} model.
For the 2-halo term we use the non-linear matter correlation
function $\xi_{\rm nl}$ scaled by the halo bias $b(M)$ of \citet{Tinker10} as
\begin{equation}
  \label{eq:2halo}
  \xi_{\rm 2h}(r\,|\,M) = b(M)\ \xi_{\rm nl}(r) \; .
\end{equation}
The non-linear matter correlation function is related
to the non-linear power spectrum $P_{\rm nl}$ as
\begin{equation}
  \label{eq:xi_nl}
  \xi_{\rm nl}(r) = \frac{1}{2\uppi}\int \mathrm{d}k\ k^2P_{\rm nl}(k)\ j_0(kr) \; ,
\end{equation}
where $j_0(kr)$ is the $0$-th spherical Bessel function of the first kind.

\subsection{Modeling systematics}
\label{sec:calibration}

Any differences between the true $\Delta\Sigma$ profiles of
cluster halos of mean mass $M$
and our analytical model for $\Delta\Sigma(M)$ from \autoref{eq:xi_hm} 
can bias the recovered weak lensing masses.  For instance, while
we chose the customary NFW profile to model the 1-halo term,
several alternatives describe the cluster
mass profiles similarly well \citep[cf.][for a recent comparison]{Umetsu14.1}.
Even within the NFW halo family, we are forced to adopt a concentration--mass relation.%
\footnote{This is a consequence of our rejection of shape measurements from $R<200$ kpc, which could otherwise allow for data-driven constraints on the concentration parameter.}
While the consequence of these choices on mass estimates are
expected to be small \citep[cf.][their section 4.3]{Hoekstra12.1}, we nonetheless
need to quantify them.

To do so, we measure the weak-lensing masses of dark matter halos in
numerical simulations using the same formalism we employ with the DES
data.  The halos are drawn from a 
$N$-body simulation of a flat $\Lambda$CDM cosmology run with the 
{\tt Gadget} code \citep{Springel05}.
The simulation uses 2.74 billion particles in a box that is 1050 Mpc $h^{-1}$ on a side.  
The matter density is $\Omega_\mathrm{m}=0.318$, 
implying that a $10^{13}\, h^{-1}\,\mathrm{M}_\odot$ halo is resolved with 
$\approx10^3$ particles. The remaining cosmological parameters are 
$H_0 = 67.04$ {km s$^{-1}$ Mpc$^{-1}$},
$\Omega_\mathrm{b}=0.049$, $\tau = 0.08$, $n_\mathrm{s} = 0.962$, and $\sigma_8 = 0.835$.
The force softening is 20 $h^{-1}$ kpc.  We discard all information below 5
softening lengths, and verified that the choice of extrapolation scheme 
for describing the correlation function below this scale does not impact 
our results. Halos are identified using the \rockstar\ halo finder 
\citep{Behroozi2013}, using a spherical overdensity mass definition of 200 times the background density.

The numerical simulation is used to construct the synthetic weak lensing signal
of dark matter halos drawn from 4 different redshift snapshots at
$z=0$, $0.25$, $0.5$ and $1$.  We split the halos in narrow mass subsets,
and compute the halo--mass correlation function with the  \citet{Landy93}
estimator from the code {\sc TreeCorr}\footnote{\url{https://github.com/rmjarvis/TreeCorr}}.
We numerically integrate the measured correlation functions to obtain 
the corresponding \dsig\ profiles as described in 
\autoref{sec:surface_density_model}. The covariance matrix of our data points 
is estimated by splitting the simulation box into 64 jackknife regions.

\begin{figure}
  \includegraphics[width=\linewidth]{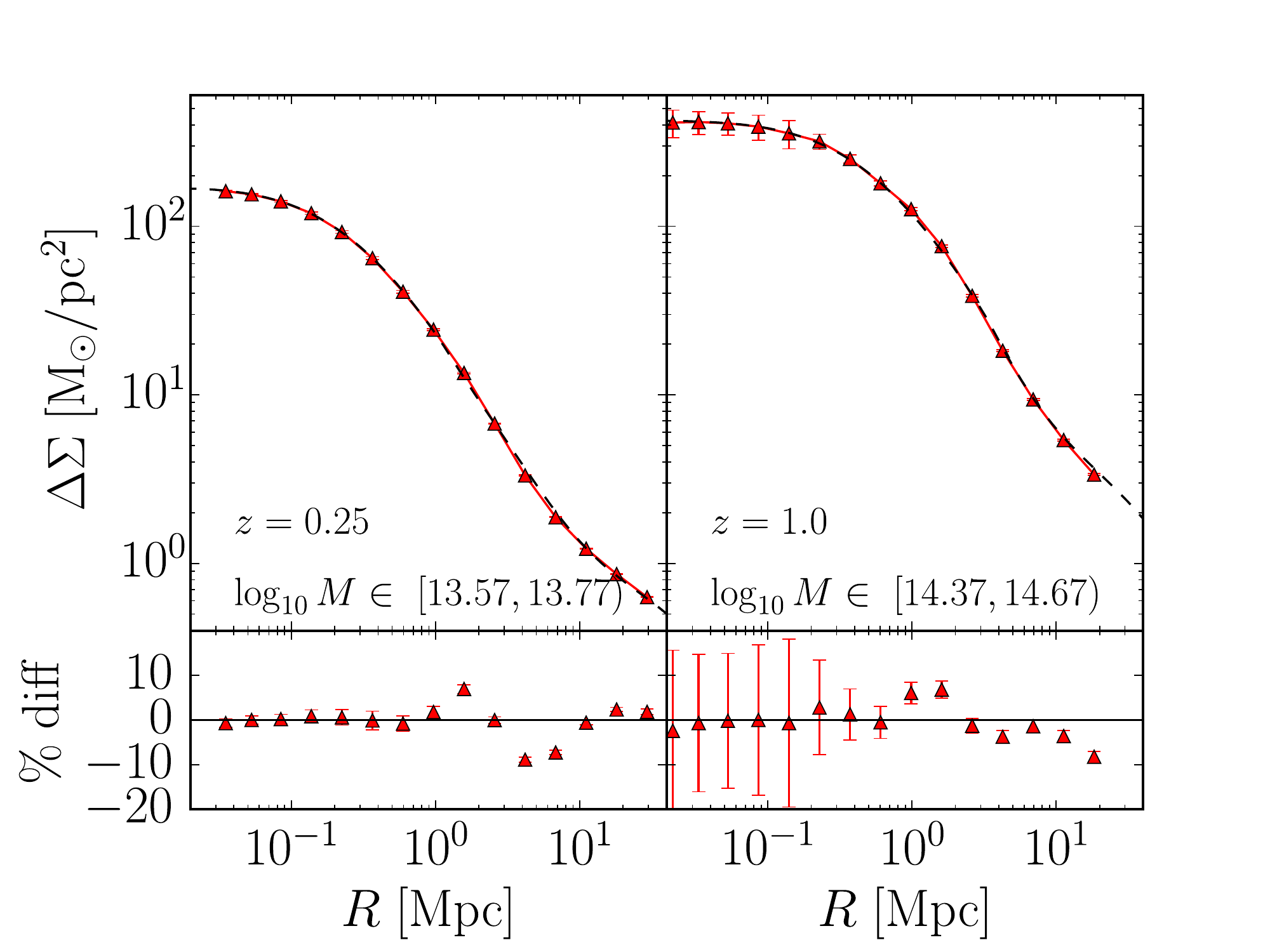}
  \caption{Simulated $\Delta\Sigma$ profile
    with error bars from jackknifing compared to the
    {\it a priori} analytical model profile from \autoref{eq:xi_hm},
    evaluated at the average mass of all halos in the associated subset.
    The two panels show the lowest redshift and lowest
    mass (\emph{left}) as well as the highest redshift and highest
    mass subset (\emph{right}), covering the mass and redshift
    range of galaxy clusters in the \redmapper\ catalog.
    Lower panels show the difference in percent between
    the analytic model and the simulated signal.
    }
  \label{fig:delta_sigma_comparison}
\end{figure}

A comparison of the simulated $\Delta\Sigma$ profiles, along with the {\it a
priori} analytical model of the $\Delta\Sigma$ profile, is shown in
\autoref{fig:delta_sigma_comparison}.  We see very small differences on small
scales, which increases to $\approx$10\% on scales of the 1-halo to 2-halo
transition. 
We tested five different mass bins at four different redshifts, and did not find a systematically
low amplitude of the model on those scales. Moreover, we note that the \citet{Tinker10} bias function
is itself only accurate at the $\pm 6\%$ level, entirely consistent with the differences seen in
\autoref{fig:delta_sigma_comparison}.
More detailed modeling \citep[e.g.][]{diemerkravtsov14} can 
reduce these deviations at the expense of additional free parameters.  
Here, we opt to empirically calibrate any biases in the recovered 
weak lensing masses due to these modest, highly localized model deviations.

\begin{figure}
  \includegraphics[width=\linewidth]{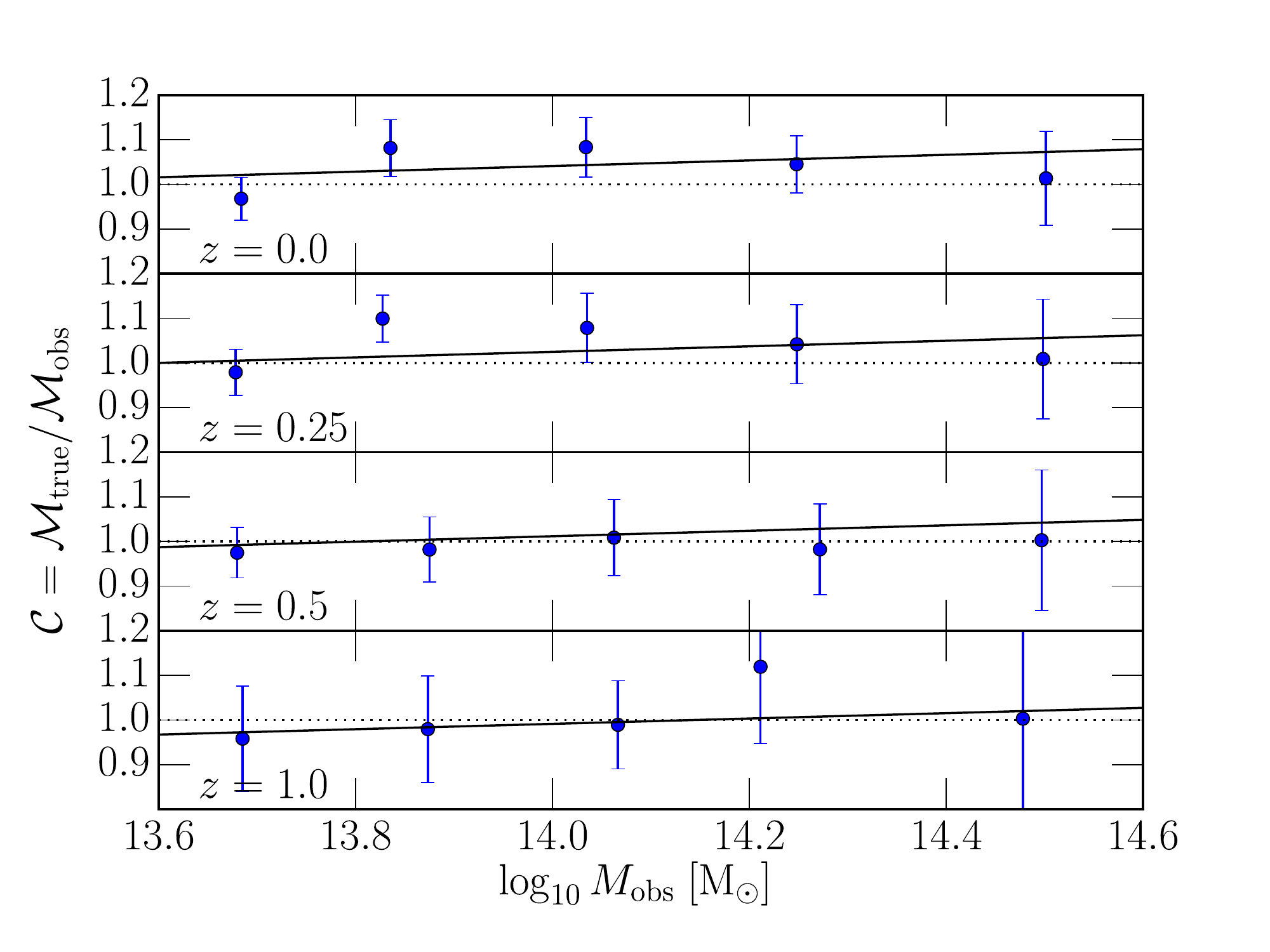}
  \caption{The mass calibration $\calC = \Mtrue\big/\Mobs$ caused by adopting 
    the analytic form of \autoref{eq:xi_hm} for the $\Delta\Sigma$ profile, as 
    a function of the observed halo mass $\Mobs$ for each simulated redshift 
    bin. The solid line is the best-fit bias model from 
    \autoref{eq:mass_bias_model}.
  }
  \label{fig:bias_plot}
\end{figure}

When doing so, we restrict ourselves to the same radial
scales employed in the weak lensing analysis, and utilize the covariance 
matrices recovered from the data. Specifically, each halo stack is fit using 
the covariance matrix of the \redmapper\ subset from \autoref{sec:covariance} 
closest in mass and redshift to the simulated cluster stack. This ensures 
that the simulated data is weighted in the same way as the observed data, 
so that any biases are appropriately calibrated. Denoting $\Mtrue$ as the 
mean mass of the simulated cluster subset and $\Mobs$ as the result of the 
corresponding lensing analysis, we find small per-cent level biases caused 
by the adopted analytical form of the $\Delta\Sigma$ profile 
(\autoref{fig:bias_plot}). We characterize the mass bias 
$\calC = \Mtrue\big/\Mobs$ as a function of the recovered weak lensing 
mass $\Mobs$ and $z$ with a power-law:
\begin{equation}
\label{eq:mass_bias_model}
\calC(\Mobs,z) = C_0\left(\frac{1+z}{1+z_0}\right)^{\alpha}\left( \frac{\Mobs}{10^{13.8}\ \msun} \right)^{\beta}
\end{equation}
with $z_0=0.5$ as pivot redshift.
The overall mass bias was found to be consistent with unity, 
with $C_0 = 1.00\pm0.02$, and mildly redshift and mass dependent, 
$\alpha = -0.071\pm0.080$ and $\beta=0.026\pm0.029$.
The redshift dependence reflects that the fit is done for each of the 
four snapshots. Given this mass calibration, the final estimator for 
the mean weak lensing mass of a cluster stack at redshift $z$ is
\begin{equation}
\label{eq:mass_corrected}
\Mobs' = \calC(\Mobs, z)\,\Mobs.
\end{equation}
We calculated any residual systematic mass errors, and found
them to be consistent with zero for all redshift and mass subsets.

\subsection{Triaxiality and projection effects}
\label{sec:tri_and_pro}

When finding galaxy clusters in photometric data, systems that are aligned
along the line of sight are typically selected with higher probability than
those otherwise oriented.  This break in isotropy must be accounted for when
inferring the mean mass density profile in a ``stacked'' analysis such as we
perform here.  Likewise, cluster selection can be affected by the presence of
other objects along the line-of-sight, which increase both the observed cluster
richness and the recovered weak lensing mass. These effects have been discussed
in a variety of works \citep{anguloetal12,whiteetal10,nohcohn12,Dietrich2014}, and they
have competing impacts on the cluster mass--richness relation.

Projection effects tend to boost the cluster richness more than the recovered
weak lensing mass.  This is easily understood: Since the weak lensing profile
$\Delta\Sigma$ is a differential measurement, any centering offset between the
projected halo and the main halo has a more severe impact on the weak lensing
signal $\Delta\Sigma$ than on the cluster richness.  The net effect of
projections is actually to somewhat reduce the mean inferred mass at given
measured richness. 

Following \citet{Simet2016}, we let $\avg{M}_0$ be the mass of a galaxy cluster
not affected by projection effects, and let $p$ be the fraction
of clusters that are affected by projections.  We model a projected cluster as a sum
of a primary halo that must have at least a mass $0.5\avg{M}_0$, and an excess
mass $\epsilon \avg{M}_0$ where $\epsilon\in [0.0,0.5]$.  We adopt a Gaussian
prior for $\epsilon$ of $\epsilon = 0.25 \pm 0.15$, so that both $\epsilon=0$ and
$\epsilon = 0.5$ are within $2\sigma$ of the central value. 
\citet{Simet2016} estimated the projection rate to be $p=10\% \pm 4\%$.

We can then write the average mass of the cluster stack as
\begin{equation}
\avg{M} = (1-p)\avg{M}_0 + p(0.5+\epsilon) \avg{M}_0
\end{equation}
The mass in the absence of projection effects is simply $\avg{M}_0$, so
to recover $\avg{M}_0$ from $\avg{M}$ we need to multiply the recovered
weak lensing masses by
\begin{equation}
\frac{\avg{M}_0}{\avg{M}} = \frac{1}{1+p(\epsilon-0.5)} = 1.02 \pm 0.02.
\end{equation}
The numerical value above is estimated from $10^4$
Monte Carlo realizations of $p$ and $\epsilon$
within the respective priors.

Unlike projection effects, the preferential alignment of 
halos along the line of sight tends to boost the recovered weak lensing mass 
at fixed measured richness relative to the true halo mass.  Using numerical simulations, \citet{Dietrich2014}
estimated this effect leads to an overestimate of cluster masses by $4.5\% \pm 1.5\%$.
This estimate can be understood as correlated scatter between optical
richness and weak lensing masses,
which leads weak lensing masses to overestimate cluster masses by an amount
$\exp(-\beta\, r\, \sigma_{\ln M|\lambda}\,\sigma_{\ln M|M_{\rm WL}})$ where $\beta\approx3$
is the slope of the halo mass function, and $r$ is the correlation coefficient
between richness and weak lensing mass.
Adopting $r\in[0,0.5]$ \citep{nohcohn12}, $\sigma_{\ln M|\lambda} = 0.25 \pm 0.05$ \citep{rozorykoff14},
and $\sigma_{\ln M|M_{\rm WL}} = 0.25 \pm 0.05$
we arrive at a correction factor $0.96\pm 0.02$, in excellent agreement with the \citet{Dietrich2014}
result. 

Put together, these two effects modify the 
recovered weak lensing masses by a multiplicative factor $0.98\pm 0.03$.%
\footnote{The balance between these two effects mildly depends on richness and redshift, while we will assume it to be constant in what follows. We will evaluate the cumulative effect of all the calibration terms in \autoref{sec:results}.}
This correction factor can be readily absorbed into the 
calibration correction parameter $C_0=1.00\pm 0.02$ described in the previous 
section, resulting in $C_0=0.98 \pm 0.04$.

\subsection{Centering correction}
\label{sec:centering}

In our calibration strategy, we have thus far assumed that we can measure the
stacked shear profile of clusters relative to the ``center'' of the halo as defined
in an $N$-body simulation.  In the simulations we used to
calibrate our weak lensing masses (\autoref{sec:calibration}), halos were found using the \rockstar\
algorithm \citep{Behroozi2013}, which identifies the halo center
as the mean position of a judiciously chosen subset of particles near the
halo's density peak. This is the point relative to which our halo--matter
correlation function is meant to be defined.

The density peak, corresponding to a ``center'' as defined
above, is expected to host a massive galaxy, which, 
in many cases, can be correctly identified from photometric data.
However, sometimes one may not be able to unambiguously determine
which of the various cluster galaxies corresponds to the
density peak.

Consequently, we adopt a model in which part of the cluster sample is correctly centered and the remaining fraction $\fmis$ is off-centered with some radial distribution $p(R_{\rm mis})$. Both $\fmis$ and the width of $p(R_{\rm mis})$ we take as free parameters in each redshift and richness subset. 
Correspondingly, we model the recovered weak lensing signal as a weighted sum of two 
independent contributions: a contribution $\Delta\Sigma$ from properly 
centered clusters,
and a contribution $\Delta\Sigma_{\rm mis}$
from miscentered galaxy clusters,
\begin{equation}
  \label{eq:miscentering_convolution}
  \Delta\Sigma_{\mathrm{model}} = (1-\fmis)\Delta\Sigma + \fmis\Delta\Sigma_{\mathrm{mis}}.
\end{equation}
When a cluster is miscentered by some radial offset $R_{\rm mis}$, the 
corresponding azimuthally averaged surface mass density is 
\citep[e.g.][]{Yang06,Johnston07}:
\begin{equation}
  \label{eq:miscentered_profile}
  \Sigma_{\rm mis}(R\,|\,R_{\mathrm{mis}}) = \int_0^{2\uppi} \frac{\mathrm{d} \theta}{2\uppi} \Sigma\left(\sqrt{R^2+R_{\mathrm{mis}}^2+2RR_{\mathrm{mis}}\cos\theta}\right).
\end{equation}

Letting $p(\Rmis)$ be the distribution of radial offsets for miscentered 
clusters, the corresponding $\Sigma_{\rm mis}$ profile is obtained by 
averaging over the ensemble,
\begin{equation}
\Sigma_{\rm mis}(R) = \int \mathrm{d}\Rmis\ p(\Rmis)\ \Sigma_{\rm mis}(R\,|\,\Rmis).
\end{equation}
\citet{Rykoff2016} modeled the distribution $p(\Rmis)$ for the DES SV 
\redmapper\ clusters with a Rayleigh distribution. The ansatz assumes that 
radial vector displacements are drawn from a two dimensional Gaussian with 
constant variance $\sigma_R$, which gives the characteristic magnitude
of the resulting radial offsets.  They further assumed that $\sigma_\mathrm{R}$
is a fraction of the cluster radius $R_\lambda$,
\begin{equation}
\sigma_\mathrm{R} = c_\mathrm{mis}\, R_\lambda.
\end{equation}
By comparing the assigned \redmapper\ cluster centers to the cluster centers 
estimated using high resolution X-ray data and SZ data from the South Pole 
Telescope, accounting also for the uncertainty of the X-ray and SZ centers, \citet{Rykoff2016} were able to place the empirical constraint
\begin{equation}
\ln c_\mathrm{mis} = -1.13 \pm 0.22
\end{equation}
and found the fraction of miscentered clusters to be
\begin{equation}
\fmis = 0.22 \pm 0.11.
\end{equation}
We will adopt these values as priors in our analysis,
though how best to do so is unclear.
One could, for instance, let the parameters $\fmis$ and $\cmis$ vary between
richness bins.  Alternatively, one could allow for redshift dependence of 
this parameters, but no richness dependence, or vice versa, or both.
We choose to allow the miscentering parameters $\fmis$ and $\cmis$ to
vary independently between subsets. Should the weak lensing data strongly 
favor models with varying $\fmis$ and $\cmis$, our
chosen parameterization should allow such trends to emerge from the data.
We further note that the priors, while determined for massive SPT clusters, conform well with analytic expectations for lower-mass clusters. We will test in \autoref{sec:results} whether our assumption of constant priors is justified.

\subsection{Boost factor model}
\label{sec:boost_factor_model}

In section \autoref{sec:boost_factors}, we noted that membership
dilution biases the recovered weak lensing
profile by a factor $1-\fcl$.  In the literature, the factor $(1-\fcl)^{-1}$
is often referred to as a boost factor or correction factor, sometimes denoted
$C(R)$,
and is used to boost the recovered profile by the appropriate amount.
While we choose to leave the data untouched---and therefore
dilute the theoretical profile rather than boost the data---we parameterize
the boost factor $\calB \equiv (1-\fcl)^{-1}$ when constructing
a model for the cluster-member contamination:
\begin{equation}
  \label{eq:boost_model}
  \calB(\lambda, z, R) = 1 + B_0 \left(\frac{\lambda}{\lambda_0}\right)^{C_\lambda}\left(\frac{1+z}{1+z_0}\right)^{D_z}\left(\frac{R}{R_0}\right)^{E_R}
\end{equation}
where $B_0$, $C_\lambda$, $D_z$, and $E_R$ are parameters in the fit.
We choose richness, redshift, and radial pivots as $\lambda_0=30$, $z_0=0.5$, 
and $R_0=500$ kpc, respectively.

We also need to address that the procedure to infer the boost factors
described in \autoref{sec:boost_factors} yielded point estimates without 
uncertainties. We account for that by modeling the associated uncertainty 
on the boost factor with the assumed form
\begin{equation}
  \label{eq:boost_error}
  \sigma_\calB(R) = \sigma_{1\ \mathrm{Mpc}}\left(\frac{1\ \mathrm{Mpc}}{R}\right)\mathrm{,}
\end{equation}
where $\sigma_{1\ \mathrm{Mpc}}$ is an unknown parameter that represents the
error at a pivot distance of $1\, \mathrm{Mpc}$.
The $1/R$ dependence is expected for logarithmically spaced radial 
bins and Poissonian errors: $\sigma_b \propto 1/\sqrt{N(R)} \propto 1/R$. 
The corresponding log-likelihood of the measured $f_{\mathrm{cl},k}$ in cluster subset $k$ given 
the parameters in \autoref{eq:boost_model} is then given by 
\begin{equation}
  \label{eq:boost_likelihood}
  \begin{split}
  \ln \lkhd (f_{\mathrm{cl},k}\,|\, B_0, C_\lambda, D_z, E_R) = &-\sum\limits_R \frac{\big((1-f_{\mathrm{cl},k}(R))^{-1}-\calB(R)\big)^2}{2\sigma_\calB^2(R)}\\
  &+\frac{1}{2}\log{\sigma_\calB^2(R)}.
  \end{split}
\end{equation}
We will constrain these parameters simultaneously from all richness and
redshift subsets. Because the boost factors also affect the $\Delta\Sigma$ 
model, we will fit it in conjunction with the lensing data to account 
for any possible degeneracies between their respective parameters.

\subsection{The complete likelihood}
\label{sec:complete_likelihood}

As we found in \autoref{sec:covariance} and \autoref{fig:cmatrix}, 
the shear profile measurements of individual richness--redshift 
subsets are nearly independent from each other. For any subset $k$ we can thus 
write the log-likelihood for a measured $\widetilde{\Delta\Sigma}$ given 
the halo mass $M$ (and the other nuisance parameters listed in \autoref{tab:modeling_parameters}) as
\begin{equation}
  \label{eq:likelihood}
  \begin{split}
  &\ln \lkhd(\Delta\Sigma_k\,|\,M_k, \dots) \propto -\frac{1}{2}\mathbf{D}_k^T\ \mathsf{C}_k^{-1} \mathbf{D}_k\mathrm{,\ where}\\
  D_{k,l} \equiv\ &\widetilde{\Delta\Sigma}_k(R_l)\ -
  \frac{A_{m,k}\ \Delta\Sigma_\mathrm{model}(R_l\,|\,M_k,f_{\mathrm{mis},k}, c_{\mathrm{mis},k})}{\calB(R_l\,|\,B_0, C_\lambda, D_z, E_R;\,k)}
    \end{split}
\end{equation}
and $\widetilde{\Delta\Sigma}(R_l)$ is the measurement in the $l$-th
radial bin from \autoref{eq:delta_sigma_est}, and $\mathsf{C}$ is the
corresponding covariance matrix from \autoref{sec:covariance}.
The factor $A_m = 1 + m - \delta$ combines the effects of shear 
($m$, \autoref{sec:sysshear})
and \photoz\ ($\delta$, \autoref{sec:sysphotoz}) systematic 
uncertainties.  Since both $m$ and $\delta$ are assigned Gaussian priors, 
the width of the prior on $A_m$ is obtained by adding
the widths of the priors on $m$ and $\delta$ in quadrature.  We arrive at
\begin{equation}
  \label{eq:multiplicative_bias}
  A_m = \begin{cases}
    1.019 \pm 0.034 & \mathrm{for}\ z\in[0.2,0.4] \\
    1.020 \pm 0.038 & \mathrm{for}\ z\in[0.4,0.6] \\
    1.044 \pm 0.039 & \mathrm{for}\ z\in[0.6,0.8].
  \end{cases}
\end{equation}
Notice that the factor $A_m$ alters the prediction 
$\Delta\Sigma_\mathrm{model}$ from \autoref{eq:miscentering_convolution},
as opposed to correcting the data. Our approach has the benefit of
preserving the covariance matrix: an alteration of the data vector would
necessarily force us to also adjust the effective covariance matrix.

The total log-likelihood for our analysis is the sum of the boost 
factor (\autoref{eq:boost_likelihood}) and weak-lensing log-likelihoods:
\begin{equation}
\label{eq:total_likelihood}
\begin{split}
\ln \lkhd =&\sum_{k=(\lambda,z)} \ln \lkhd_k\ \ \mathrm{with}\\
\ln \lkhd_k\equiv&\ln \lkhd(\Delta\Sigma_k\,|\,M_k,A_{m,k},f_{\mathrm{mis},k}, c_{\mathrm{mis},k}, B_0, C_\lambda, D_z, E_R;\,k)\ +\\
 &\ln \lkhd(f_{\mathrm{cl},k}\,|\,B_0, C_\lambda, D_z, E_R;\,k).
 \end{split}
 \end{equation}
It is thus apparent that we seek to constrain independent subset masses 
$M_k$ and global boost factor parameters, the latter constrained by their 
effect on the $\Delta\Sigma$ profile as well as independent measurements 
of $f_\mathrm{cl}$.

\subsection{Stacked cluster masses}
\label{sec:results}

The likelihood is sampled using the package \textit{emcee}%
\footnote{\url{http://dan.iel.fm/emcee}} \citep{Foreman13} that
allows a parallelized exploration of the parameter space.
We use 20 walkers with 10,000 steps each, and discard
the first 3,000 steps as burn-in. We test whether the chains have converged
first with an independent run of only 5,000 steps per walker,
which produces nearly identical results. 
%
The chains of single walkers become uncorrelated (with a correlation coefficient $|r|<0.1$) after about 23 steps, which is much shorter than the length of each chain.  The resulting number of independent draws for all walkers is $\approx 6000$.
We therefore believe that the likelihood
has been exhaustively explored and that our inference results are robust.

The complete list of model parameters as well as their corresponding
priors and posteriors are summarized in \autoref{tab:modeling_parameters}.

After determining the best fit masses $M$ for each  cluster subset,
we apply the calibration correction described in \autoref{sec:calibration} 
to the recorded chains.
Specifically, for each point in the chain, we randomly sample the 
mass calibration factor $\calC(M,z)$ from its posteriors, and 
replace the mass parameter value with $\Mobs'$ from \autoref{eq:mass_corrected}.
Such a postponed correction is valid because $\calC$ is independent from
other parameters in the chain, and results in an updated chain that 
incorporates the mass calibration and its corresponding uncertainties.

\begin{table}
\setlength{\tabcolsep}{.4em}
  \caption{Parameters entering $\lkhd(\Delta\Sigma)$
    (\autoref{eq:likelihood}) and $\lkhd(f_\mathrm{cl})$
    (\autoref{eq:boost_likelihood}). Flat priors are specified
    with limits in square brackets, Gaussian priors with means $\pm$
    standard deviations. The posteriors 
    are given as the mean and
    symmetrized 68\% confidence intervals. For the top half, the posteriors
    depend on the cluster subset, so we list the uncertainties as average 
    confidence intervals followed by the scatter between subsets.}
   \label{tab:modeling_parameters}
     \begin{tabular}{llll}
       Parameter & Description & Prior & Posterior \\ \hline
       $\log_{10}M$ & Halo mass & $[12.0,16.0]$ & \autoref{tab:posterior_masses} \\
       $\ln c_{\rm mis}$ & Miscentering offset & $-1.13\pm0.22$ & $-1.06 \pm 0.22 \pm 0.05$ \\
       $f_{\rm mis}$ & Miscentered fraction & $0.22\pm0.11$ & $0.24 \pm 0.10 \pm 0.05$ \\
       $A_{m}$&Shape \& \photoz\ bias & \autoref{eq:multiplicative_bias} & $1.026 \pm 0.037 \pm 0.013$ \\
       \hline
       $\log_{10}B_0$ & Boost magnitude & $[-4.0,0.0]$ & $-1.399\pm0.040$ \\
       $C_\lambda$ & Richness scaling & $[0.25,1.5]$ & $0.920\pm0.106$ \\
       $D_z$ & Redshift scaling & $[-10,10]$ & $-4.00\pm0.79$ \\
       $E_R$ & Radial scaling & $[-1.5,1.0]$ & $-0.98\pm0.09$ \\
     \end{tabular}
\end{table}

\begin{figure}
  \includegraphics[width=\linewidth]{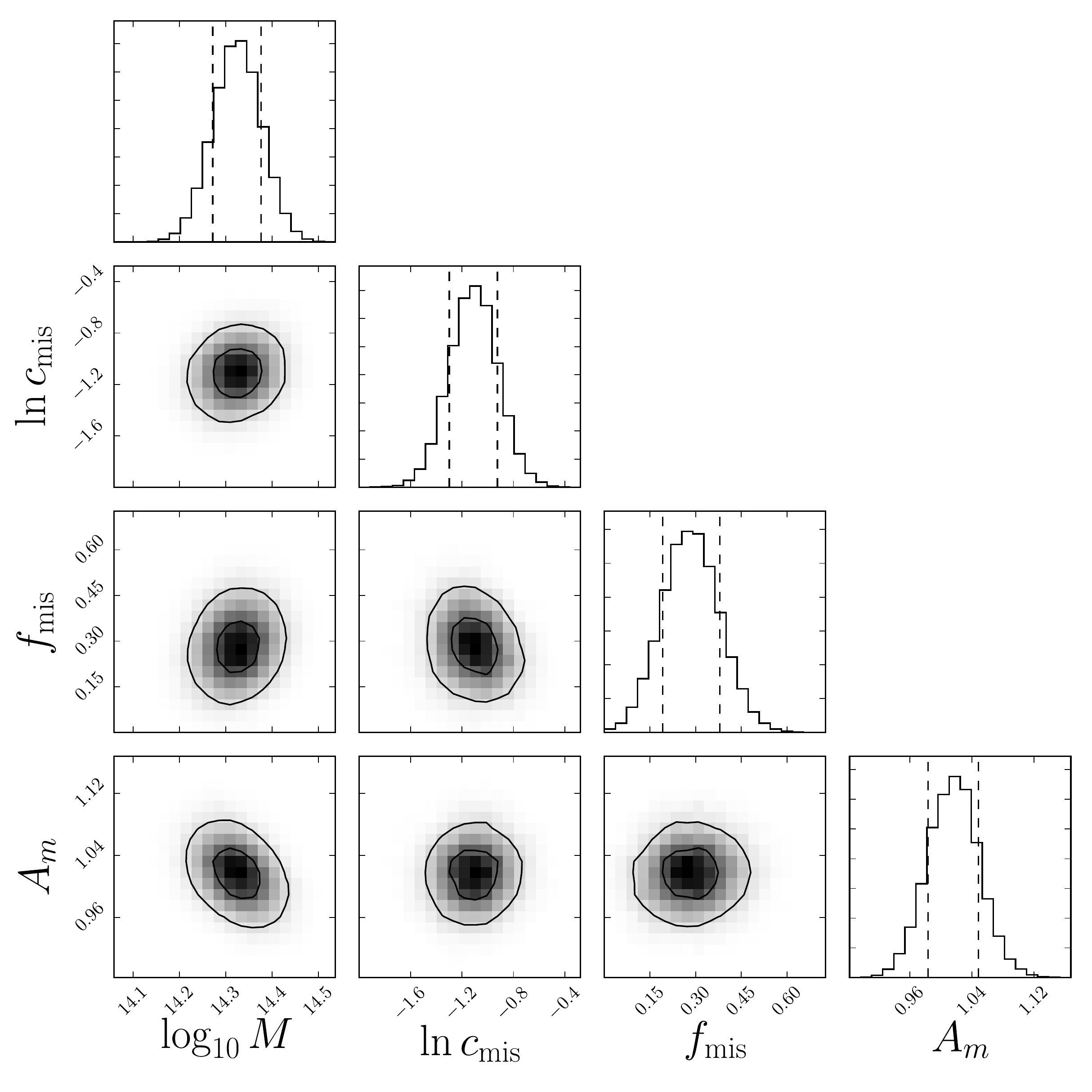}
  \caption{Parameters of $\Delta\Sigma_\mathrm{model}$ for
    the $z\in[0.2;0.4)$, $\lambda\in[20,35)$ subset. Contours
    denote the 68\% and 95\% confidence areas; dashed lines in the 
    1D histograms refer to 68\% confidence intervals.}
  \label{fig:topcorner}
\end{figure}

\begin{figure}
  \includegraphics[width=\linewidth]{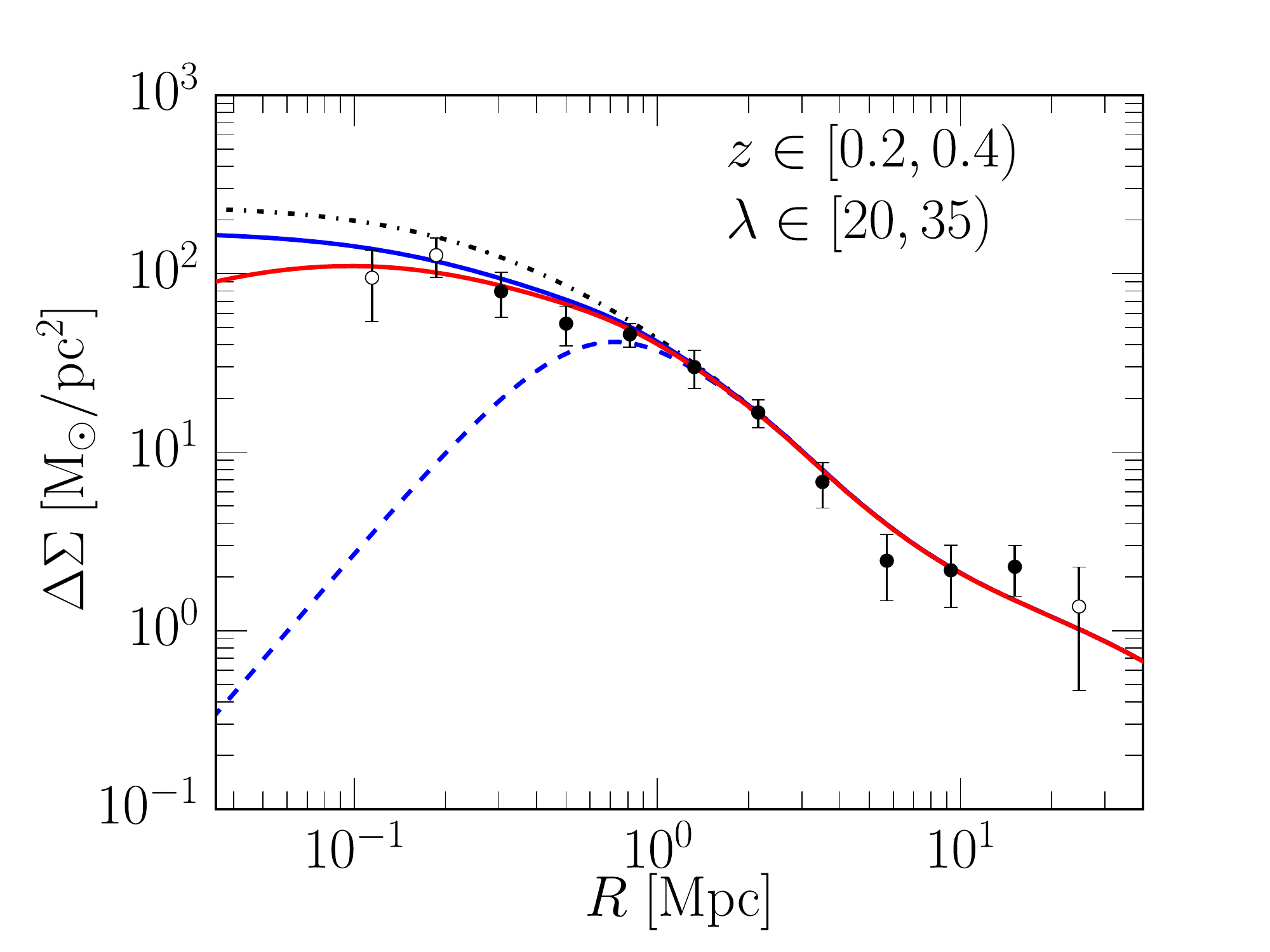}
  \caption{Fit with all components of $\Delta\Sigma_\mathrm{model}$ 
    and $\mathcal{B}$ to
    the cluster subset with $z\in[0.2;0.4)$ and $\lambda\in[20,35)$.
    The analytical model for the perfectly centered lensing signal 
    $\Delta\Sigma$ from \autoref{eq:xi_hm} is shown as 
    \emph{dash-dotted black curve}. 
    The \emph{solid blue curve} includes the effects of miscentering.
    The \emph{solid red curve} additionally includes the effects of 
    cluster member contamination from \autoref{eq:boost_model} and 
    corresponds to the model that is fit to the data, shown as 
    best-fit curves in \autoref{fig:DeltaSigma}.
    We exclude data at $R< 200\ \mathrm{kpc}$ (\emph{open markers}) to avoid 
    problems due to crowded field photometry and large boost-factor corrections in the cluster core.
    The \emph{dashed blue curve} corresponds to the miscentered 
    profile $\Delta\Sigma_\mathrm{mis}$.}
  \label{fig:best_fit_example}
\end{figure}

We run four variants of the final likelihood evaluation to 
quantify statistical and systematic uncertainties in our analysis.
\begin{itemize}[label=\textbullet, leftmargin=*]
\item \verb|Full|: All systematic parameters (modeling bias parameters, 
triaxiality and projection, shape and \photoz\ systematics, boost factors, and miscentering)
are allowed to vary within their respective priors.  This constitutes our fiducial analysis.
\item \verb|FixedAm|: All systematic parameters are allowed to vary within their priors, except we set $A_m=1$
to determine the influence of only the combined shape and \photoz\ uncertainties.
\item \verb|Fixed|: All systematic parameter priors are set to $\delta$-functions at their central values
to estimate the statistical uncertainties.
\item \verb|NonLinear|: Identical to \verb|Full| but with a modified data vector $\Delta\Sigma\rightarrow(1-\kappa)\Delta\Sigma$,
where the convergence $\kappa$ was measured from the simulated mass profiles with the same subsets in mass and redshift as in \autoref{sec:calibration}, to approximate the non-linearity bias on the weak-shear estimator in \autoref{eq:reduced_shear}.
\end{itemize}
The results of the \verb|Full| likelihood evaluation for the parameters
of $\Delta\Sigma_\mathrm{model}$ under the priors from 
\autoref{tab:modeling_parameters} are shown in \autoref{fig:topcorner} 
for the example richness--redshift subset of $z\in[0.2,0.4)$ and 
$\lambda\in[20,35)$. The corresponding lensing data and best-fit lensing 
profile are shown in \autoref{fig:best_fit_example}, where we also single 
out the impact of miscentering and boost factors.
The best-fit models for all \redmapper\ cluster subsets are 
over-plotted on top of our weak lensing data in \autoref{fig:DeltaSigma}.

As we can see from the posteriors in \autoref{tab:modeling_parameters},
the boost factors amount to an correction $B_0\approx-4\%$ at the pivot
values of $\lambda_0=30$, $z_0=0.5$, $R_0=\mathrm{500\ kpc}$ and are consistent
with linear scaling in $\lambda$ and $1/R$ radial scaling, both 
expected from the number density of cluster member galaxies in the 
inner part of an NFW halo.

\begin{table*}
  \caption{The calibrated marginalized posterior masses for each richness--redshift subset.
    Masses are given as $\log_{10}[M_{200\mathrm{m}}]$ in units of $M_\odot$,
    uncertainties denote symmetrized 68\% confidence intervals marginalized
    over all other parameters.
    We first list the contributions from the statistical, then systematic uncertainties.
    In case direct matching to NFW halo results is desired, these can be obtained by reverting the calibration of \autoref{eq:mass_bias_model}.}
  \label{tab:posterior_masses}
    \begin{tabular}{llll}
      $\lambda$ & $z\in[0.2,0.4)$ &$z\in[0.4,0.6)$ & $z\in[0.6,0.8)$ \\ \hline
	$[5,10)$ & $13.300\pm 0.095 \pm 0.025$ & $13.457\pm 0.056 \pm 0.025$ & $13.442\pm 0.141 \pm 0.027$ \\
	$[10,14)$ & $13.758\pm 0.115 \pm 0.011$ & $13.520\pm 0.158 \pm 0.015$ & $13.637\pm 0.152 \pm 0.033$ \\
	$[14,20)$ & $14.034\pm 0.087 \pm 0.029$ & $13.962\pm 0.088 \pm 0.024$ & $14.096\pm 0.149 \pm 0.019$ \\
	$[20,35)$ & $14.324\pm 0.065 \pm 0.018$ & $14.297\pm 0.080 \pm 0.018$ & $14.114\pm 0.122 \pm 0.024$ \\
	$[35,180)$ & $14.592\pm 0.070 \pm 0.017$ & $14.619\pm 0.080 \pm 0.020$ & $14.664\pm 0.089 \pm 0.021$ \\
    \end{tabular}
\end{table*}

Posteriors on the miscentering parameters are only weakly constrained by our data.  We find a weak correlation of $M$ with $f_\mathrm{mis}$ (and none with $c_\mathrm{mis}$) in most cluster subsets, but there is no apparent trend of either miscentering parameter with richness or redshift.
While tighter constraints could in principle be found by combining the lensing measurement with angular clustering of the putative centers and member galaxies \citep{vanUitert2016}, we find that, at least for now, the uncertainties in mass due to miscentering are sub-dominant to statistical and other systematic uncertainties (cf.~\autoref{tab:budget}).
We postpone further investigation of cluster miscentering to forthcoming works with larger DES data volumes.

To quantify statistical and systematic uncertainties of the fiducial analysis, we then perform the \verb|Fixed| likelihood evaluation. 
We determine the systematic contribution to the fiducial uncertainty as the difference of uncertainties in quadrature between the \verb|Full| and the \verb|Fixed| run. In \autoref{tab:posterior_masses} 
we list central values and uncertainties for all subsets, and split the latter into statistical and systematic contributions.

Finally, to estimate the impact of the weak-shear assumption $\mathbf{g}\approx\bgamma$ in \autoref{eq:reduced_shear}, we perform the \verb|NonLinear| run, where we applied the first-order correction $\Delta\Sigma\rightarrow(1-\kappa)\Delta\Sigma$.
We find a mild overestimation of the masses from the \verb|Full| run by $\approx 2\%$ for the highest richness subsets; all other subsets are 
affected at the sub-percent level.
Given the uncertainties in the current analysis, we will ignore this bias and its potential impact on the richness--mass relation. For future analyses we will adopt correction schemes to suppress the non-linearity bias of the shear estimator \citep[e.g.][]{Seitz97.1,Johnston07.2}.

\section{The mass--richness--redshift relation}
\label{sec:mass_richness_relation}

We characterize the mass--richness relation of the DES SV \redmapper\
galaxy clusters as
\begin{equation}
  \label{eq:mass_richness_redshift}
  \calM(\lambda,z) \equiv \avg{M\,|\,\lambda,z} = M_0\left(\frac{\lambda}{\lambda_0}\right)^{F_\lambda}
  \left(\frac{1+z}{1+z_0}\right)^{G_z},
\end{equation}
where $M_0$, $F_\lambda$, and $G_z$ are parameters of the model with
pivot values $\lambda_0 = 30$ and $z_0 = 0.5$. We note the important distinction between
$M$, the mass of a halo, itself a random variable, and $\calM$, the expectation value of
that random variable.
For each cluster subset $k$, the expectation value is given by
\begin{equation}
  \label{eq:weighted_mass}
  \calM_k = \frac{\sum_{j\in k}W_j\calM(\lambda_j,z_j)}{\sum_{j\in k} W_j}.
\end{equation}
The weights $W_j$ of individual clusters $j$ in subset $k$ differ from unity for two reasons: 1) the lensing weight of each
lens--source pair depends on the cluster's redshift, and 2) lower redshift clusters have more
sources in any given radial bin because a fixed physical radial bin of a low redshift cluster
subtends a larger angle in the sky than the same bin does for a higher redshift cluster.
We estimate the weight $W_j$ as the sum of weights $w_{j,i}$ of all lens--source pairs around cluster $j$ 
(given in \autoref{eq:wdeltasigma}) over the radial range $0.3$~Mpc to $3$~Mpc.  The choice
of radial range has a sub-percent impact on our results.  

There is one subtle effect that remains unaccounted for in the above formula: in practice,
we do not stack cluster masses.  Rather, we stack the density profiles $\Delta\Sigma$.  Using
our analytic model for $\Delta\Sigma$, we can readily estimate the logarithmic dependence
on mass of the density profile $\Delta\Sigma$ at any given radius $R$
\begin{equation}
  \label{eq:gamma_derivative}
  \Gamma(R) = \frac{d\ln \Delta\Sigma(R\,|\,M)}{d\ln M}.
\end{equation}
This logarithmic slope varies from $\approx 0.5$ in the innermost regions of the density profile we utilize to $\approx 1.0$
on the outskirts, with a typical value of $\approx 0.75$ over a broad range of scales.  For specificity, from here on
we compute $\Gamma$ within the radius $R_{200\rm m}$, for which we find $\Gamma=0.74 \pm 0.01$ within the
range of masses and redshifts probed in this work.  Given that we stack $\Delta\Sigma \propto M^\Gamma$, and that
we then recovered the mean $\Delta\Sigma$ profile and turn it into a mass, i.e. $M\propto \Delta\Sigma^{1/\Gamma}$,
we arrive at the appropriately weighted mean mass for each subset
\begin{equation}
  \label{eq:weighted_mass2}
  \calM_k = \left(\frac{\sum_{j\in k}W_j\calM(\lambda_j,z_j)^\Gamma}{\sum_{j\in k} W_j}\right)^{1/\Gamma}.
\end{equation}
For $\Gamma=1$ this reduces to \autoref{eq:weighted_mass}.
We compare the predictions of this model to the mass 
$M_k$ of subset $k$ measured from lensing after marginalizing over all other 
parameters (cf. \autoref{tab:posterior_masses}). Since we find the 
posterior probability density of $M_k$ to be best described as log-normal
(as expected from \citealt{Stanek10}), 
we can express the log-likelihood of the mass--richness relation parameters as
\begin{equation}
  \label{eq:mass_richness_likelihood}
  \ln \lkhd(\Mobs\,|\,M_0,F_\lambda,G_z) \propto - \frac{1}{2} (\Delta \log M)^T\ \mathsf{C}_M^{-1}\ (\Delta \log M)\mathrm{,}  
\end{equation}
where $\mathsf{C}_M$ is the covariance matrix of the logarithm of 
the inferred masses $M_k$, which we shall derive below, and
\begin{equation}
\Delta \log M_k = \log M_k - \log \calM_k.
\end{equation}

The covariance matrix $\mathsf{C}_{M}$, however, does not have a trivial form.
While we have found the statistical 
errors of lensing masses from the different richness and redshift subsets to 
be nearly uncorrelated (as discussed in \autoref{sec:covariance}), 
the systematic uncertainties are necessarily correlated.  The primary
source of calibration uncertainty was found to be model bias due to using
a simple exponential disk to model all source galaxies.  While the bulge
fraction is certainly a function of redshift, the lensing kernel is broad
enough that we expect the overall calibration error to be similar
for all bins in lens redshift and richness.
Likewise, \photoz\ systematics impact the mass measurements of all clusters. 
We therefore consider the recovered mass $M_k$ and its 
prediction $\calM_k$, and write uncertainties in $\log M_k$ as a sum of an 
uncorrelated component $\mu_k$ and a correlated component $\nu_k$:
\begin{equation}
\log M_k = \log \calM_k
+ \mu_k + \nu_k.
\end{equation}
The diagonal elements of the covariance matrix are given by
\begin{equation}
\mathsf{C}_{M,kk} = \avg{\mu_k^2} + \avg{\nu_k^2}\mathrm{,}
\end{equation}
which is equal to the posterior uncertainties of the \verb|Full| run in \autoref{tab:posterior_masses}.
We then evaluate the \verb|FixedAm| run, which sets the potentially correlated parameter $A_{m,k}=1\ \forall k$ and thereby
enforces $\nu_k=0$, so that the recovered errors correspond to the uncorrelated statistical noise component $\avg{\mu_k^2}$ only.
Subtracting the two errors, we arrive at the correlated noise
$\avg{\nu_k^2}$.  

Having solved for the correlated noise $\avg{\nu_k^2}$ for each 
subset $k$, we compute the full covariance matrix $\mathsf{C}_{M,ij}$
by assuming the \photoz\ and shear systematic uncertainties
are fully correlated between different cluster subsets.
This assumption is conservative 
as the uncertainties are not reducible 
by ``averaging out'' the impact of these systematics across different subsets.
We thus set
\begin{equation}
\label{eq:C_M_correlated}
\mathsf{C}_{M,ij} = \delta_{ij}\avg{\mu_i^2} + \avg{\nu_i \nu_j} \\ = \delta_{ij} \avg{\mu_i^2} + \left[ \avg{\nu_i^2}\avg{\nu_j^2} \right]^{1/2}
\end{equation}
as the covariance matrix we employ in \autoref{eq:mass_richness_likelihood}.

\begin{table}
\setlength{\tabcolsep}{.6em}
  \caption{Parameters of the $M$--$\lambda$--$z$ relation from 
    \autoref{eq:mass_richness_redshift} with their flat priors and 
    resulting posteriors. The mass is defined as $M_{200\mathrm{m}}$ 
    in units of $\msun$.
    Uncertainties denote symmetrized 68\% confidence intervals and are split
    into statistical (first) and systematic (second).}
  \label{tab:mass_richness_parameters}
    \begin{tabular}{llll}
      Parameter & Description & Prior & Posterior \\ \hline
      $\log_{10}M_0$ & Mass pivot & $[12.0,16.0]$ &  $14.371 \pm 0.040 \pm 0.022$ \\
      $F_\lambda$ & Richness scaling & $[-10,10]$ & $1.12 \pm 0.20 \pm 0.06$ \\
      $G_z$ & Redshift scaling & $[-20,20]$ & $0.18 \pm 0.75 \pm 0.24$ \\
    \end{tabular}
\end{table}

\begin{figure}
  \includegraphics[width=\linewidth]{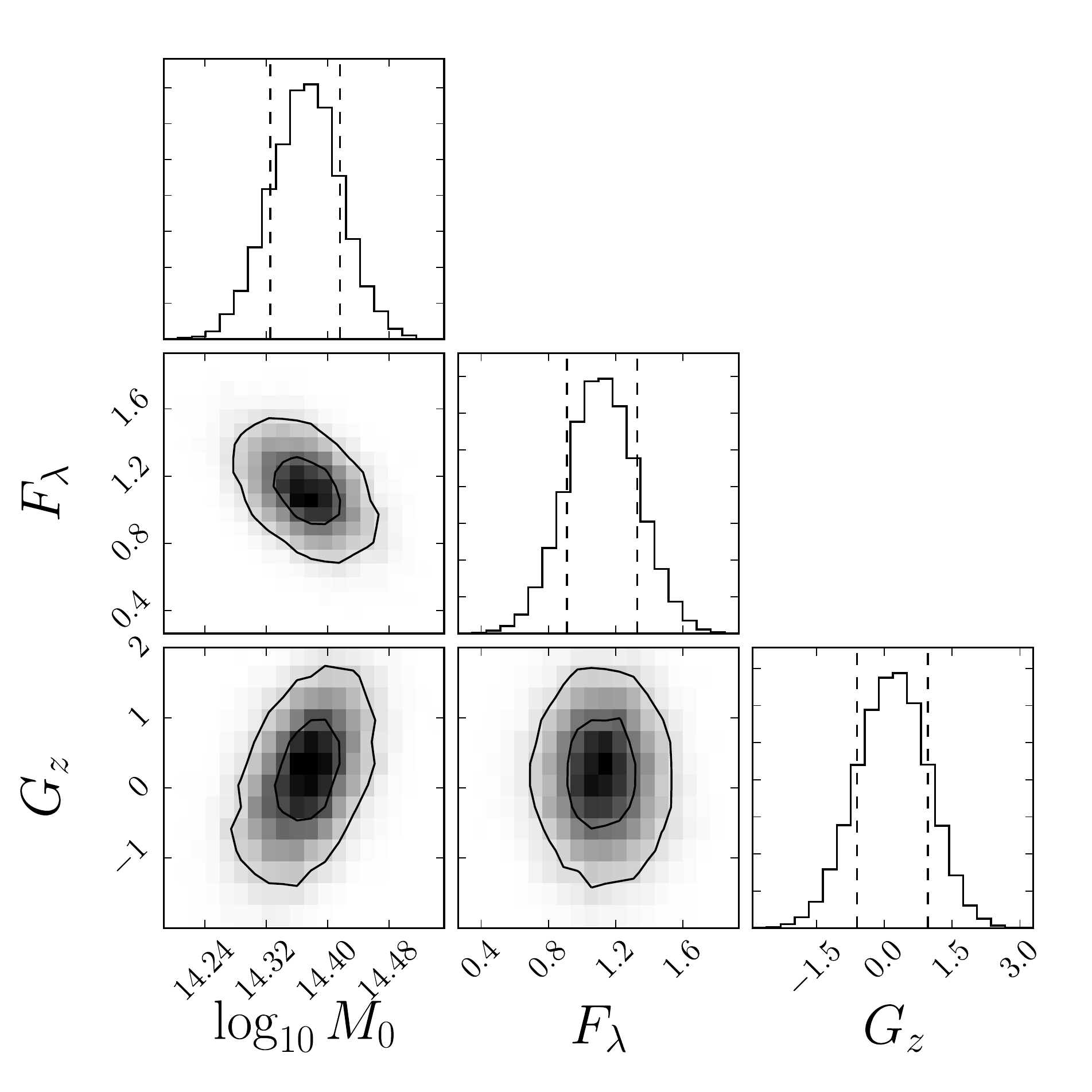}
  \caption{Parameters of the $M-\lambda-z$ relation.
  Contours denote the 68\% and 95\% confidence areas from the \texttt{Full} run; dashed lines in
  the 1D histograms refer to the 68\% confidence intervals.}
  \label{fig:mr_corner}
\end{figure}

Note that with this scheme, we explicitly enforce correlated uncertainties of shear and \photoz\ systematics only,
whereas other systematics are considered independent across subsets.
Independent systematics will tend to ``average out'' across bins, reducing their impact on the
uncertainty in the amplitude of the scaling relation, while increasing their impact on the slope
of the scaling relation.  In our case, assuming non-\photoz\ and non-shear systematics
are independent across bins reduces their effect on the amplitude of the mass--richness
relation from 2\% on each independent subset to $\approx 1\%$ 
(see \autoref{tab:budget} for our systematics error budget).

As before, we use \textit{emcee} to sample the likelihood
of the mass--richness relation parameters as constrained
from the DES SV data.  Our fit is restricted to subsets with $\lambda\geq 20$
to ensure that galaxy clusters can be unambiguously identified with prominent dark matter
halos.\footnote{The richness threshold $\lambda\geq 20$ has been adopted in
all \redmapper\ works since its original publication as a sufficiently conservative
cut to ensure clean cluster samples. As reference, the typical richness uncertainty due to
random projections for a clusters with $\lambda = 5$ is $\pm 1.5$, therefore for a $\lambda=20$
system to not be associated with at least one halo of richness $\lambda=5$ would constitute a
$10\sigma$ fluctuation.}
The best-fit parameters are summarized in \autoref{tab:mass_richness_parameters}, and the 
corresponding confidence contours are shown in \autoref{fig:mr_corner}.

We repeat the analysis using the statistical errors from the \verb|Fixed| run.
That is, we constrain the mass--richness using statistical errors only.
The central values of the resulting parameters are nearly 
identical to the parameters we inferred
when marginalizing over all systematic uncertainties.  
The difference in quadrature between the two uncertainties is reported as the 
systematic uncertainty in the recovered mass--richness--redshift relation
in \autoref{tab:mass_richness_parameters}.  

We also carry out the analysis with an extreme value $\Gamma=1$ in 
\autoref{eq:weighted_mass2} to address the concern that our treatment does not fully 
capture the variation in $\Gamma$ across all the scales being utilized.
We find that it changes the recovered amplitude of the mass--richness relation 
by $\Delta \log M_0 = 0.003$, which is clearly subdominant compared to the other 
uncertainties in the analysis.

Our results imply that the mean mass of galaxy clusters of
richness $\lambda = 30$ at redshift $z=0.5$ is
$\log \calM = 14.371$ $\pm 0.040\ \mathrm{(stat)}$ $\pm 0.022\ \mathrm{(sys)}$,
with a richness scaling that is slightly steeper than linear and no strong redshift evolution.
This corresponds to a 10.5\% calibration (9.2\% statistical, 5.1\% systematic)
of the amplitude of the mass--richness relation.

\begin{figure}
  \includegraphics[width=\linewidth]{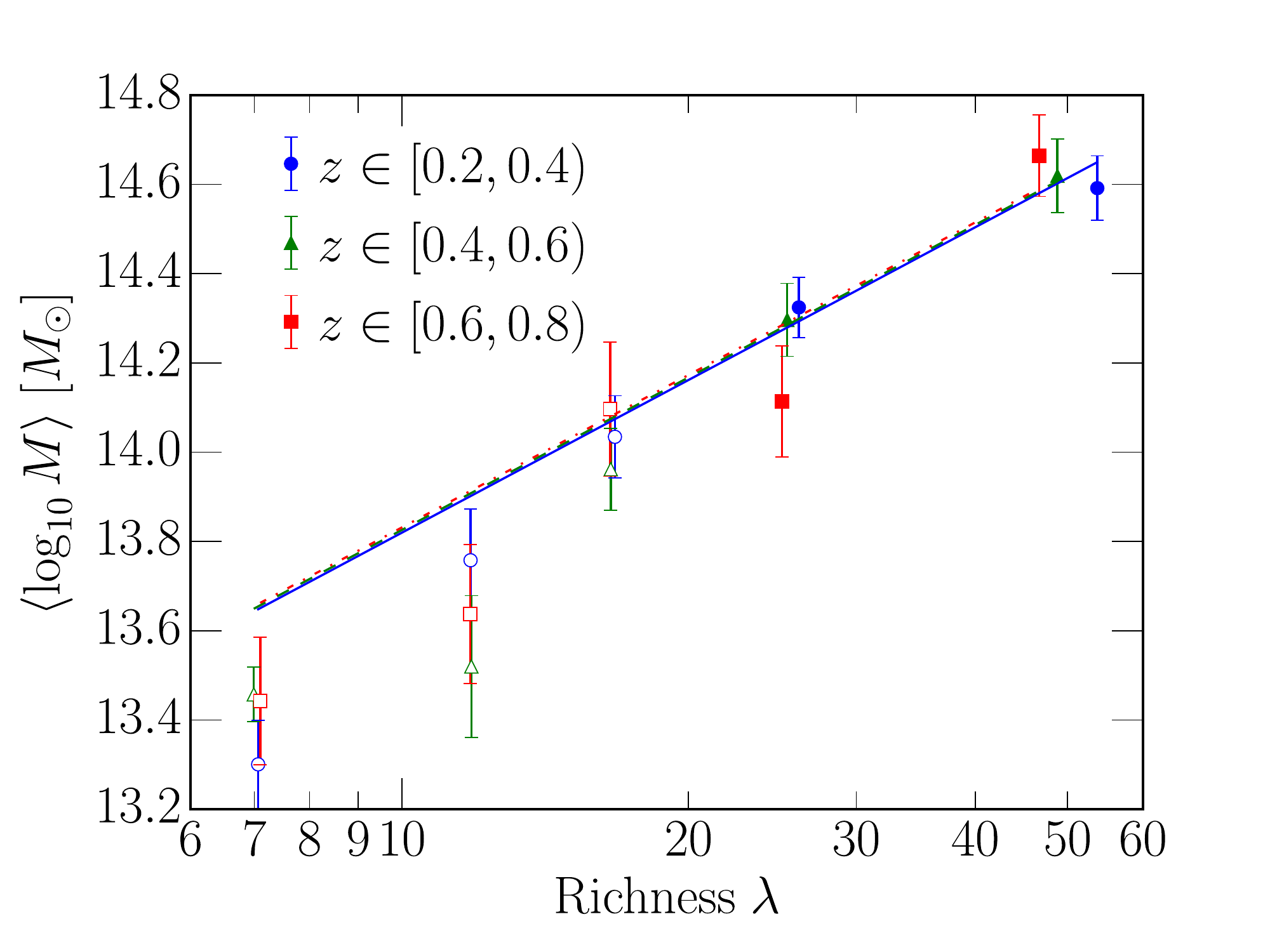}
  \caption{Best fit model for $M(\lambda,z)$. Only subsets with
    richnesses $\lambda\geq20$ (\emph{filled markers}) are fit. Points are placed at the
    mean richness of their subset.}
  \label{fig:mass_richness}
\end{figure}

The current statistical uncertainties reflect the small area of the DES SV data set;
we expect a similar analysis of the full survey will yield $\approx 1\%$ uncertainty
after five years of operations.
These results highlight the need for substantial improvement in systematic 
error control within a short time frame.
  
Like \citet{Simet2016}, we expect the dominant
systematic uncertainty in our analysis to stem from 
shear and \photoz\ systematics.  
We confirm this by repeating the analysis with the \verb|FixedAm| run, which accounts for all systematics
{\it except} multiplicative shear bias and \photoz\ systematics.
As expected, the recovered posterior distributions are significantly 
narrower than the posteriors that include shear and \photoz\ systematics.  
We find that those two systematics alone contribute
78\% of the systematics-sourced variance.  The remaining 22\% of the systematics
variance is almost entirely due to projections, triaxiality, and modeling systematics.   

\autoref{fig:mass_richness} compares the best-fit mass--richness
relation to the stacked cluster masses $M_j$. Our model presents an excellent fit 
for the considered range of $\lambda \geq 20$, and extrapolates well down to the $\lambda \geq 14$
cluster sample.  Clusters with even lower richnesses fall significantly below our prediction, likely reflecting
contamination from line-of-sight overdensities.  Our results 
are consistent with the expectation that a richness threshold $\lambda \geq 20$ is a conservative choice for 
cluster scaling-relation studies.

\section{Comparison to results in the literature}
\label{sec:comparisons}

We now compare our calibration of the mass--richness relation to 
results in the literature, based on SPT SZE measurements of DES \redmapper\ 
clusters (\autoref{sec:saro2015})
and weak lensing by SDSS \redmapper\ clusters (\autoref{sec:simet}). We also compare our 
systematic uncertainties to previous cluster lensing studies (\autoref{sec:discusssys}).

\subsection{Comparison to Saro et al. (2015)}
\label{sec:saro2015}

\citet[][hereafter S15]{Saro2015} provided the first, indirect calibration of the richness--mass relation
of DES \redmapper\ clusters by cross-matching the SV cluster
catalog to the SPT cluster catalog of \citet{Bleem2015}.
To do so, S15 first assumed a cosmology identical to our
fiducial cosmology (flat $\Lambda$CDM with $\Omega_\mathrm{m}=0.3$, $h=0.7$,
with $\sigma_8=0.8$), then determined the best-fit relation between the 
SZE signal-to-noise ratio $\xi$ and the cluster mass by abundance matching,
i.e.~by comparing the predicted number of clusters above the observable threshold to a prediction 
based on the mass function and observable--mass relation.
This allowed them to transfer the $\xi-M$ relation from the SPT
clusters to the $\lambda-M$ relation of DES \redmapper\ clusters thanks to the well-understood SPT
cluster selection function.

To compare our results to S15, we directly sample the scaling relation parameters from the S15 chains.
We then need to account for a key difference between our work and that of S15, 
namely that we constrain the mass--richness relation, while the latter constrains 
the richness--mass relation. One can use the approach of \citet{Evrard2014} to
transform between the two. We refer the reader to that work for details, and
simply note that the conversion requires a correction for the scatter in mass at 
fixed value of the observable. The correction itself depends on the first and second 
logarithmic derivatives of the halo mass function, which we compute using the 
\citet{Tinker2008} halo mass function evaluated at $z=0.6$, the pivot redshift in the S15 relation.

\begin{figure}
  \includegraphics[width=\linewidth]{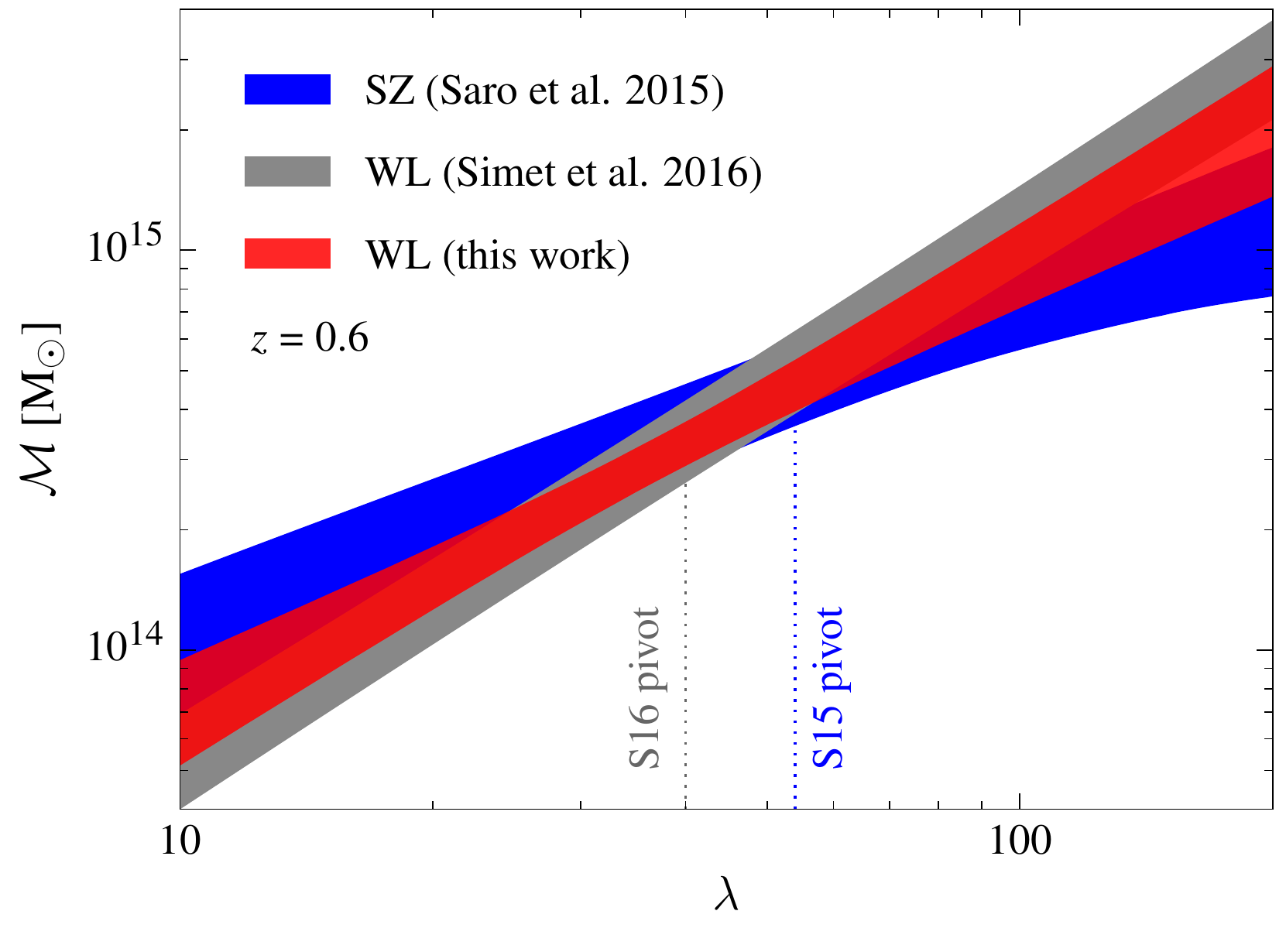}
  \caption{68\% confidence intervals for the mean cluster mass $M_{\rm{200m}}$ as function
  of cluster richness $\lambda$ at $z=0.6$ as constrained by the analyses 
  of \citet[][\emph{blue}]{Saro2015},
  \citet[][\emph{gray}]{Simet2016}, and this work (\emph{red}). The \citet{Simet2016} relation has
  been transported from its pivot redshift of 0.2 to 0.6 using our best-fit redshift-evolution model.
 }
\label{fig:compare}
\end{figure}

\autoref{fig:compare} compares the recovered scaling relation from S15 to our 
results, after also converting
the S15 masses from $M_{\rm{500c}}$ in units of $\hinv\ \msun$ to $M_{\rm{200m}}$ in units of $\msun$, 
using the method described by \citet{Hu2003} and the \citet{Bhattacharya2013} concentration--mass relation.
Essentially identical results are obtained using the \citet{DiemerKravtsov15} relation.
Although visually the slope of the mass--richness relation
from S15 is shallower than ours, the difference is not significant: the two are consistent
at the $1.3\sigma$ level.   The amplitude of the scaling relations are in nearly perfect agreement
agreement: the S15 mean mass at their pivot richness of $\lambda=54$ 
is $\log \avg{M\,|\,\lambda=54}=14.67 \pm 0.11$,
compared to $\log \avg{M\,|\,\lambda=54} = 14.66 \pm 0.06$ in our analysis.

\subsection{Comparison to SDSS}
\label{sec:simet}

Several studies have calibrated the mass--richness relation for SDSS \redmapper\ 
clusters \citep[S16 hereafter]{Li2015,Miyatake2016,Farahi2016,Baxter2016,Simet2016}. 
All of them have been found to be consistent with each
other.  While the \citet{Li2015} calibration is the most precise (i.e.
it has the smallest reported errors), it does not include a detailed
analysis of the sources of systematic uncertainty.
The measurements by \citet{Farahi2016} and \citet{Baxter2016}
are useful systematics cross-checks, but are currently less precise than
those from S16.  Given the statistical consistency between the various methods, 
and the fact that S16 is the most precise after accounting for systematic uncertainties,
we focus on the comparison with S16.

S16 find that the mean mass of SDSS \redmapper\ clusters of richness $\lambda=40$
is $\log \avg{M\,|\,\lambda=40} = 14.344 \pm 0.021\ \rm{(stat)} \pm\ 0.023\ \rm{(sys)}$ at 
a reported pivot of $z=0.2$.  In that work, masses were measured as $M_{200\rm m}$ in units of $\hinv\ \msun$.  
At that richness and redshift, our results, converted to their units, correspond
to $\log \avg{M\,|\,\lambda=40} = 14.339 \pm 0.070$, fully consistent with---but not as precise as---their result.
The slopes of the mass--richness relations are also in good agreement,
with a difference of $0.9\sigma$.

We note that DES \redmapper\ clusters in the SV data are somewhat
more abundant than SDSS \redmapper\ clusters at the same richness, although the difference does not
appear to be statistically significant \citep[their Figure 6]{Rykoff2016}.
Nevertheless, a difference in abundance could in principle signify a systematic difference
in the richness estimator $\lambda$ between DES and SDSS, resulting in the same galaxy clusters
being assigned different richnesses when observed in DES than when observed in SDSS. 
Conversely, clusters of the same observed richness $\lambda$, either from DES or from SDSS,
would, on average, have somewhat different masses. The difference in cluster abundance
between the two surveys could thus imply that
DES SV clusters are about 7\% less massive than equally rich SDSS clusters.%
\footnote{The ratio between the comoving space density of galaxy clusters in DES vs SDSS
for clusters with richness $\lambda \geq 20$ over the redshift range $z\in [0.2,0.3]$, where both DES and SDSS are complete, is 0.83 \citep{Rykoff2016}. 
If we assume that this ratio corresponds to different effective mass thresholds in the two surveys, we can use abundance matching
to estimate the corresponding systematic mass difference between the two surveys.
}
Correcting for this effect, our prediction for the SDSS mass calibration would be 
$\log \avg{M\,|\,\lambda=40} = 14.373 \pm 0.053$, still in excellent agreement with the SDSS result.
We emphasize that this effect does not affect the uncertainty in the mass calibration of the DES 
SV galaxy clusters; it is relevant only for the comparison of our mass calibration to that of S16.
Whether the difference in number density is indeed due to a lower mass threshold in DES or merely a statistical fluctuation 
can be directly tested with future DES \redmapper\ catalogs with larger area and partial SDSS overlap.

\subsection{Comparison of systematic uncertainties to other weak-lensing cluster mass calibrations}

\label{sec:discusssys}

Over the last two years, several collaborations have published results of extensive observational campaigns
designed to calibrate cluster mass--observable relations.  These include the
Weighing the Giants \citep[WtG,][]{WtGI}, the Canadian Cluster Comparison Project \citep[CCCP,][]{Hoekstra2015}, and the 
Local Cluster Substructure Survey \citep[LoCuSS,][]{Okabe2015}. 
We also include the weak-lensing mass measurements from 
the Cluster Lensing and Supernova Survey with Hubble 
\citep[CLASH,][]{CLASH1,Umetsu2016} in this discussion, although the calibration of mass--observable relations 
is not their explicit goal.  
A direct comparison of our results with these works is not possible
because we do not estimate individual cluster masses, and because of a lack of overlap between the clusters samples.
Nevertheless, a discussion on the respective treatment of systematic uncertainties is warranted.
For completeness, we also discuss the systematic error budget of S16.

We focus here on the two main sources of systematic uncertainty, multiplicative shear bias ($m$, cf. \autoref{eq:shear_bias}) and \photoz\ calibration ($\delta$, cf. \autoref{eq:delta}).
For the former, each of the collaborations relied on simulations to calibrate their shear biases and to estimate the corresponding uncertainties.
\wtg\ employed calibrations from STEP2 \citep{STEP2} and state shear bias uncertainties of 3\% \citep{Applegate2014}.
\clash\ used STEP2 supplemented by similar, custom simulations from \citet{Oguri2012}, and find 5\% uncertainty in $m$ \citep{Umetsu2016}.
\cccp\ used {\sc galsim} \citep{Rowe2014} to simulate analytic galaxy profiles, and quote 2\% shear calibration uncertainty.
\locuss\ utilized two different sets of image simulations, one using the software package {\sc SkyMaker} \citep{Bertin2009}
and another using the software package {\sc Shera} \citep{Mandelbaum2012}, and state 3\% uncertainty in $m$.
S16 quote 3.5\% top-hat systematic uncertainty, roughly equivalent to 2.0\% Gaussian uncertainty,
based on {\sc Shera} simulations.

The shear systematics error budget is comparable across all works.
We caution, however, that our shear systematic
is larger than what we would have estimated from the GREAT-DES simulations \citep[][their section 6.1]{Jarvis2016} alone.
Our systematic error estimate comes from the comparison of two independent source catalogs, \imshape\ and
\ngmix, each of which we expected to have a multiplicative shear bias $|m|\leq 0.03$.  Nevertheless, a detailed
comparison of the two revealed a systematic uncertainty $|m|\leq 0.05$. We believe this increased systematic uncertainty 
reflects differences between our simulated images and our data.  

There is a larger discrepancy between our work and how others have 
estimated photometric redshift systematics, which we find to be the second most important source
of systematic uncertainty in our analysis.
We emphasize that throughout this section we use the term ``\photoz\ systematics'' to denote
uncertainties in $\avg{\Sigma_{\rm{crit}}^{-1}}$ caused exclusively by biased performance
of the photometric redshift estimator, e.g. due to insufficient template sets or priors and cosmic variance limiting 
the precision of the calibration. 
This systematic is distinct from cluster-member dilution, and any associated amelioration techniques such as the 
color cuts employed in many studies (including \wtg, \locuss, and \clash).  We make this distinction because
we find membership dilution to be a sub-dominant effect.

As with estimation of shear calibration systematics, our photometric redshift systematic error is estimated by comparing
several independently produced \photoz\ catalogs.  The corresponding systematic uncertainty in the weak lensing
signal ranges from 1.7\% for low-redshift clusters to 2.5\% for high-redshift clusters (cf. \autoref{eq:delta_prior}). 
This matches well the systematic
uncertainty estimated directly from weighted spectroscopic validation data sets \citep{BonnettPhotoz2015}.

\wtg, \cccp, \locuss, and \clash\ approach the problem of estimating \photozs\ of the selected sources differently.
\wtg\ \citep{Kelly2014} and \clash\ use the template-fitting code \bpz\ (one of the codes we also employ) for determining photometric redshifts from 5-band photometric data.
Analyses with fewer imaging bands infer
indirect \photoz\ estimates by matching the respective source samples according to magnitude (\cccp), or color and magnitude (\wtg, for their clusters with fewer than 5 photometric bands), or $N$-nearest neighbors in color-magnitude space (\locuss)  
to a high-quality reference sample with excellent \photoz\ accuracy.
All of these studies make use of the deep multi-band photometry in the COSMOS field, and derive the reference \photoz\ estimates with the template codes {\sc LePhare} (\wtg\  and \clash, with the public catalog of \citealt{Ilbert2009}, \locuss\ with \citealt{Ilbert2013}) or {\sc eazy} \citep[for \cccp]{Brammer2008}.

The quoted systematic uncertainties in the lensing amplitudes from possible biases in the inferred redshift distribution
vary from work to work, but range from 1\% or less (\wtg\ with 5-band photometry, \locuss, \clash) to as high as 
4\% (\wtg\ with 2-bands) and 8\% for high-redshift \cccp\ clusters.  The latter two reflect cosmic variance
for single-color and/or magnitude selected source galaxies.  Our 4-band \photoz\ error of $\approx 3\%$ in the mass 
is somewhat intermediate between the two extremes above.  It is an accurate reflection of the uncertainty spanned 
by different \photoz\ codes with 4-band photometry, and in that sense it represent a {\it minimum} systematics floor
due to algorithmic choices.  It is, however, conceivable that our own \photoz\ error is somewhat underestimated
if the variation across different \photoz\ codes is less than the impact of sample variance on the machine learning \photoz\
codes.  In \citet{BonnettPhotoz2015} we found the spread in \photoz\ codes to be comparable to observed uncertainties 
relative to $p(z)$ estimates of independent spectroscopic testing samples, we therefore believe our procedure to be fair.
Nevertheless, future studies of this source of systematic uncertainty are highly desirable.

In this context, we consider the work by \citet{Nakajima12} as an instructive baseline.  
Because the SDSS shape catalog is restricted to
sufficiently bright galaxies, one can construct spectroscopic galaxy samples 
that are exactly representative of  the weak-lensing source samples.
In this ideal scenario, the accuracy with which \photoz\ biases can be controlled is limited only by the size
of these spectroscopic calibration sample, corresponding to a 3\% precision for SDSS 
clusters.
In contrast to the SDSS analysis, sufficiently large, representative spectroscopic samples are currently unavailable at the depth of DES,
and this lack is even more relevant at the depths of typical \wtg, \cccp, \locuss, and \clash\ images.
While some alleviation of this systematic naturally occurs with increasing separation of lenses and sources,
it is difficult to imagine a superior scenario for \photoz\ bias calibration than that of \citet{Nakajima12}.
This suggests that controlling \photoz\ systematics at the 2\% level or better with DES or other
upcoming large scale surveys will be difficult unless 
new methods for handling spectroscopic incompleteness are developed, 
and/or we are able to employ alternate redshift estimation techniques, e.g. the cross-correlation method of \citet{newman08}.

A reasonable test for the field of cluster lensing, as a whole, is the comparison of cluster masses across different works.  
From Table 6 in \citet{Okabe2015}, we see that, depending
on the precise mass definition, \locuss\ differs from \cccp\ by 5\%, and from \wtg\ by 12\%. 
\cccp\ reports their masses are in excellent agreement with \wtg\, with a mean offset of $\approx 2\%$,
though they caution that \cccp\ differs from the \wtg\ effort in how they convert shear profiles to masses.
In particular, when \cccp\ follows the \wtg\ prescription for estimating cluster masses, their recovered masses are lower than \wtg\
by 8\%.
The \citet{Umetsu2016} \clash\ analysis is consistent with \wtg\ and \locuss, yet finds marginally higher masses than \cccp\ by
$16\% \pm 10\%$.
These comparisons suggests that at present discrepancies between different groups remain at the $\approx 10\%$
level. Assuming equal parts statistical and systematic errors and ignoring possible overlaps of the cluster samples, 
systematic errors budgets of $\approx 7\%$ appear adequate for the current state of
the field, and are 
in reasonable agreement with corresponding estimates quoted in these works.

\section{Summary and Conclusions}
\label{sec:summary}

We measured the stacked weak-lensing
signal of \redmapper\ clusters in the DES SV data.  The clusters
were split into 15 non-overlapping richness and redshift subsets with
$\lambda \geq 5$ and $0.2 \leq z \leq 0.8$, and the mean mass of
each cluster stack was estimated from a model that accounts for:
\begin{itemize}[label=\textbullet, leftmargin=*]
\item Shear measurement systematics (\autoref{sec:sysshear}),
\item Dilution of the source sample by cluster members (\autoref{sec:boost_factors}),
\item Source photometric redshift uncertainties (\autoref{sec:sysphotoz}),
\item Analytical modeling systematics (\autoref{sec:calibration}),
\item Triaxiality \& projection effects (\autoref{sec:tri_and_pro}),
\item Cluster miscentering (\autoref{sec:centering}).
\end{itemize}
The set of masses were in turn used to determine the cluster mass--richness relation,
as parameterized according to \autoref{eq:mass_richness_redshift}.
The entire analysis was performed with a blinded shear
catalog, with a blinding factor between 0.9 and 1, which corresponds
to 13\% uniform uncertainty on the amplitude of the mass--richness
relation, comparable to a Gaussian uncertainty of about $8\%$.
By comparison, the total (statistical plus systematic) uncertainty in the amplitude of the mass--richness relation is $10.5\%$,
which implies that the blinding factor was marginally sufficient to avoid confirmation bias.

The mean cluster mass for clusters at our pivot richness of $\lambda=30$
and pivot redshift of $z=0.5$ is
\begin{equation}
M_0 = \left[2.35 \pm 0.22 \pm 0.12\right] \cdot 10^{14}\ \msun.
\end{equation}
The best-fit slope $F_\lambda$ for the mass--richness relation is
\begin{equation}
F_\lambda = 1.12 \pm 0.20 \pm 0.06,
\end{equation}
while the best-fit redshift-evolution parameter $G_z$ is
\begin{equation}
G_z = 0.18 \pm 0.75 \pm 0.24.
\end{equation}
The results are summarized in \autoref{tab:posterior_masses} and \autoref{tab:mass_richness_parameters}.

We compared our inferred mass--richness relation for \redmapper\ clusters
to that of \citet[S15]{Saro2015}.  The two works are in nearly perfect agreement
with regards to the amplitude of the mass--richness relation.  
Our recovered slope is steeper,
but the constraints from the two works are statistically consistent at $1.2\sigma$.
Since the S15 work used the abundance of SPT galaxy clusters and an $\Omega_m=0.3$
and $\sigma_8=0.8$ flat $\Lambda$CDM cosmology, the excellent agreement between
our work and that of S15 suggests that this cosmology is likely to provide a good fit
to the abundance of \redmapper\ clusters.  An analysis of the abundance of \redmapper\
clusters is in preparation.

We also compared our results to the weak lensing mass calibration of SDSS \redmapper\ 
clusters presented in \citet[S16]{Simet2016}. Our results are in excellent agreement with
those of S16, both in the amplitude and slope.  The excellent agreement between the
two works is encouraging given the difference in imaging and methods between
the two works. We caution, however, that the uncertainty in the amplitude of the mass--richness
relation of \redmapper\ clusters at the S16 pivot redshift is $\approx 16\%$, so the current
agreement is not sufficient to test the consistency of the two works within the reported
systematics uncertainties (roughly 5\% each at their respective pivot redshifts).

While our cluster mass calibration is currently statistics-limited, the situation will quickly change with the advent
of more DES data.  As \autoref{tab:budget} clearly shows, the systematic error budget is strongly 
dominated by calibration uncertainties of the multiplicative shear bias and 
the \photoz\ performance. The latter become increasingly important
at higher redshifts. 

In this work, we have assumed that \photoz\ and shear measurement systematics are perfectly 
correlated across all richness and redshift subsets.  
While it is clear that these systematics must be correlated at some level, it is not obvious to what degree.
Our conservative assumption (in terms of the 
amplitude of the mass--richness relation) of perfect correlation means that there is no ``averaging'' of these systematics
across subsets. 
All other systematics were assumed to be uncorrelated.
If we enforced perfect correlation on these systematics as well, our error budget would increase from 5.1\% to 6.1\%.
In either case, the cumulative systematic uncertainty is comparable to reported values from other weak lensing analyses
(i.e. \wtg, \cccp, \locuss, \clash, and S16).

\begin{table}
\caption{Systematic error budget on the amplitude of the mass--richness relation as measured with
the DES SV data. The first two sources are taken as perfectly correlated between source subsets, while the next three are assumed to be independent (in brackets we list their impact if considered perfectly correlated as well). In case systematic errors are determined for the amplitude of the lensing signal $\Delta\Sigma$, we use the approximation $M\propto \Delta\Sigma^{4/3}$ (see discussion of \autoref{eq:gamma_derivative}). 
}
\label{tab:budget}
\setlength{\tabcolsep}{1.5em}
\begin{tabular}{llcc}
Source of systematic & Amplitude uncertainty \\
\hline
Shear measurement &  4\% \\
Photometric redshifts & 3\%\\ 
Modeling systematics & 1\%\ (2\%) \\
Cluster triaxiality & 1\%\ (2\%) \\
Line-of-sight projections & 1\%\ (2\%) \\
Membership dilution + miscentering & $\leq 1\%$ \\
\hline
{\bf Total} & {\bf 5.1\%} (6.1\%)
\end{tabular}
\end{table}

An insight from our analysis is that systematic differences from independently
produced shape and photometric redshift catalogs can reveal systematics that
would otherwise go unnoticed. For the multiplicative shear calibration, the
differences between the shear catalogs may partly be due to differences between
DES data and the GREAT-DES simulations used to calibrate one of the pipelines.
The discrepant behavior of the \photoz\ estimators is less well understood, but
likely due to a strong reliance on a comparison to the COSMOS field, whose
small size and redshift distribution is not representative of the DES survey.

With the rapid increase in observed DES area, future mass-calibration analyses using
the same shear and \photoz\ methods would be systematics-limited. Efforts aimed at improving control over shear measurement and photometric
redshift uncertainties are thus paramount. 
In the short term, shear measurement systematics can be reduced through improvements in the
imaging simulations used to calibrate shear biases. More substantial gains will rely
on shear estimation algorithms that are less sensitive to measurement noise and assumptions of PSF and galaxy properties 
\citep[e.g.][]{Bernstein14.1}, or by determining the calibrations directly from observational data, without reference
to external simulations. 

For the long-standing issue of \photoz\ calibration, a concerted spectroscopic effort is now urgently needed to accurately characterize the mapping between galaxy colors and $n(z)$ \citep[following e.g.][]{Masters15.1}, as well as a thorough understanding of the limitations of current \photoz\ estimation schemes.
Alternatively, cross-correlation \photoz\ methods as proposed by \citet{newman08} may provide another calibration tool,
but it remains to be seen whether they can achieve the $\approx 1\%$
level accuracy required to render this source of error sub-dominant. 
We thus anticipate that the mass calibration of galaxy clusters identified
in DES Year-1 data will be limited by \photoz\ uncertainties.
Assuming we can reduce our shear measurement uncertainties to the 1\% level (1.3\% on the mass),
the next DES results should allow us to constrain the amplitude of the mass--richness relation to 
$\approx 5\%$.
Any improvements in photometric redshift uncertainties will further reduce this error, directly benefitting 
forthcoming cluster cosmology studies.

\section*{Acknowledgments}

Support for DG was provided by NASA through the Einstein
Fellowship Program, grant PF5-160138. TM and ER are supported by
DOE grant DE-SC0015975.
TNV was supported
by the DAAD (Deutscher Akademischer Austauschdienst), the SFB-Transregio 33 `The Dark Universe' by the Deutsche
Forschungsgemeinschaft (DFG) and the DFG cluster of excellence
`Origin and Structure of the Universe'. ES is supported by
DOE grant DE-AC02-98CH10886.  ER acknowledges support
from the Sloan Foundation, grant FG-2016-6443.

We are grateful for the extraordinary contributions of our CTIO
colleagues and the DECam Construction, Commissioning and Science Verification
teams in achieving the excellent instrument and telescope
conditions that have made this work possible.  The success of this project also
relies critically on the expertise and dedication of the DES Data Management group.

Funding for the DES Projects has been provided by the U.S.
Department of Energy, the U.S. National Science Foundation,
the Ministry of Science and Education of Spain,
the Science and Technology Facilities Council of the United
Kingdom, the Higher Education Funding Council for England,
the National Center for Supercomputing
Applications at the University of Illinois at Urbana-Champaign,
the Kavli Institute of Cosmological Physics at the University of Chicago,
the Center for Cosmology and Astro-Particle Physics at the Ohio State University,
the Mitchell Institute for Fundamental Physics and
Astronomy at Texas A\&M University, Financiadora de Estudos e Projetos,
Funda{\c c}{\~a}o Carlos Chagas Filho de Amparo {\`a} Pesquisa do
Estado do Rio de Janeiro, Conselho Nacional de
Desenvolvimento Cient{\'i}fico e Tecnol{\'o}gico and
the Minist{\'e}rio da Ci{\^e}ncia, Tecnologia e Inova{\c c}{\~a}o,
the Deutsche Forschungsgemeinschaft and the Collaborating
Institutions in the Dark Energy Survey.

The Collaborating Institutions are Argonne National Laboratory,
the University of California at Santa Cruz, the University of
Cambridge, Centro de Investigaciones Energ{\'e}ticas,
Medioambientales y Tecnol{\'o}gicas-Madrid, the University of
Chicago, University College London, the DES-Brazil Consortium,
the University of Edinburgh,
the Eidgen{\"o}ssische Technische Hochschule (ETH) Z{\"u}rich,
Fermi National Accelerator Laboratory, the University of Illinois
at Urbana-Champaign, the Institut de Ci{\`e}ncies de l'Espai (IEEC/CSIC),
the Institut de F{\'i}sica d'Altes Energies, Lawrence Berkeley
National Laboratory, the Ludwig-Maximilians Universit{\"a}t
M{\"u}nchen and the associated Excellence Cluster Universe,
the University of Michigan, the National Optical Astronomy Observatory,
the University of Nottingham, The Ohio State University, the
University of Pennsylvania, the University of Portsmouth,
SLAC National Accelerator Laboratory, Stanford University, the
University of Sussex, Texas A\&M University, and the OzDES Membership Consortium.

The DES data management system is supported by the National
Science Foundation under Grant Number AST-1138766.
The DES participants from Spanish institutions are partially supported
by MINECO under grants AYA2012-39559, ESP2013-48274, FPA2013-47986,
and Centro de Excelencia Severo Ochoa SEV-2012-0234 and SEV-2012-0249.
Research leading to these results has received funding from the
European Research Council under the European Union's Seventh
Framework Programme (FP7/2007-2013) including ERC grants
agreements 240672, 291329, 306478.

This work used simulations and computations performed using computational resources at SLAC and at NERSC.

\section*{Affiliations}
{\small
\begin{enumerate}[label=$^{\arabic*}\,$, leftmargin=*, widest=10, align=left, itemsep=1pt]
\item Department of Astrophysical Sciences, Princeton University, Peyton Hall, Princeton, NJ 08544, USA
\item Kavli Institute for Particle Astrophysics \& Cosmology, P. O. Box 2450, Stanford University, Stanford, CA 94305, USA
\item SLAC National Accelerator Laboratory, Menlo Park, CA 94025, USA
\item Department of Physics, University of Arizona, Tucson, AZ 85721, USA
\item Max Planck Institute for Extraterrestrial Physics, Giessenbachstrasse, 85748 Garching, Germany
\item Universit\"ats-Sternwarte, Fakult\"at f\"ur Physik, Ludwig-Maximilians Universit\"at M\"unchen, Scheinerstr. 1, 81679 M\"unchen, Germany
\item Brookhaven National Laboratory, Bldg 510, Upton, NY 11973, USA
\item Department of Physics, ETH Zurich, Wolfgang-Pauli-Strasse 16, CH-8093 Zurich, Switzerland
\item Department of Physics, Stanford University, 382 Via Pueblo Mall, Stanford, CA 94305, USA
\item Fermi National Accelerator Laboratory, P. O. Box 500, Batavia, IL 60510, USA
\item Kavli Institute for Cosmological Physics, University of Chicago, Chicago, IL 60637, USA
\item Department of Physics and Astronomy, Pevensey Building, University of Sussex, Brighton, BN1 9QH, UK
\item Jodrell Bank Center for Astrophysics, School of Physics and Astronomy, University of Manchester, Oxford Road, Manchester, M13 9PL, UK
\item Department of Physics and Astronomy, University of Pennsylvania, Philadelphia, PA 19104, USA
\item Excellence Cluster Universe, Boltzmannstr.\ 2, 85748 Garching, Germany
\item Faculty of Physics, Ludwig-Maximilians-Universit\"at, Scheinerstr. 1, 81679 Munich, Germany
\item Department of Physics \& Astronomy, University College London, Gower Street, London, WC1E 6BT, UK
\item Department of Physics, University of California, Santa Cruz, CA 95064, USA
\item Computer Science and Mathematics Division, Oak Ridge National Laboratory, Oak Ridge, TN 37831, USA
\item Center for Cosmology and Astro-Particle Physics, The Ohio State University, Columbus, OH 43210, USA
\item Department of Physics, The Ohio State University, Columbus, OH 43210, USA
\item Institut de F\'{\i}sica d'Altes Energies (IFAE), The Barcelona Institute of Science and Technology, Campus UAB, 08193 Bellaterra (Barcelona) Spain
\item Jet Propulsion Laboratory, California Institute of Technology, 4800 Oak Grove Dr., Pasadena, CA 91109, USA
\item Cerro Tololo Inter-American Observatory, National Optical Astronomy Observatory, Casilla 603, La Serena, Chile
\item Department of Physics and Electronics, Rhodes University, PO Box 94, Grahamstown, 6140, South Africa
\item CNRS, UMR 7095, Institut d'Astrophysique de Paris, F-75014, Paris, France
\item Sorbonne Universit\'es, UPMC Univ Paris 06, UMR 7095, Institut d'Astrophysique de Paris, F-75014, Paris, France
\item Laborat\'orio Interinstitucional de e-Astronomia - LIneA, Rua Gal. Jos\'e Cristino 77, Rio de Janeiro, RJ - 20921-400, Brazil
\item Observat\'orio Nacional, Rua Gal. Jos\'e Cristino 77, Rio de Janeiro, RJ - 20921-400, Brazil
\item Department of Astronomy, University of Illinois, 1002 W. Green Street, Urbana, IL 61801, USA
\item National Center for Supercomputing Applications, 1205 West Clark St., Urbana, IL 61801, USA
\item Institut de Ci\`encies de l'Espai, IEEC-CSIC, Campus UAB, Carrer de Can Magrans, s/n,  08193 Bellaterra, Barcelona, Spain
\item Institute of Cosmology \& Gravitation, University of Portsmouth, Portsmouth, PO1 3FX, UK
\item School of Physics and Astronomy, University of Southampton,  Southampton, SO17 1BJ, UK
\item Department of Physics, IIT Hyderabad, Kandi, Telangana 502285, India
\item Instituto de F\'isica Te\'orica IFT-UAM/CSIC, Universidad Aut\'onoma de Madrid, Cantoblanco 28049, Madrid, Spain
\item Department of Physics, University of Michigan, Ann Arbor, MI 48109, USA
\item Astronomy Department, University of Washington, Box 351580, Seattle, WA 98195, USA
\item Australian Astronomical Observatory, North Ryde, NSW 2113, Australia
\item Departamento de F\'{\i}sica Matem\'atica,  Instituto de F\'{\i}sica, Universidade de S\~ao Paulo,  CP 66318, CEP 05314-970, S\~ao Paulo, SP,  Brazil
\item Department of Astronomy, The Ohio State University, Columbus, OH 43210, USA
\item Department of Astronomy, University of Michigan, Ann Arbor, MI 48109, USA
\item Instituci\'o Catalana de Recerca i Estudis Avan\c{c}ats, E-08010 Barcelona, Spain
\item Centro de Investigaciones Energ\'eticas, Medioambientales y Tecnol\'ogicas (CIEMAT), Madrid, Spain
\item Universidade Federal do ABC, Centro de Ci\^encias Naturais e Humanas, Av. dos Estados, 5001, Santo Andr\'e, SP, Brazil, 09210-580
\end{enumerate}
}

\bibliographystyle{mn2e_adsurl}
\bibliography{apj-jour,astroref}

\label{lastpage}
\end{document}